\documentclass[intlimits,twoside,a4paper]{article}

\usepackage{amsmath,amssymb}
\usepackage{graphicx}
\usepackage{wrapfig}

\usepackage[T2A]{fontenc}
\usepackage[cp1251]{inputenc}

\usepackage[eqsecnum]{cmpj2}

\issue{2012}{15}{2}{23604}
\doinumber{10.5488/CMP.15.23604}
\title[Equation of state in cluster-forming systems]%
{Effect of mesoscopic fluctuations on equation of state in cluster-forming systems%
}
\author[A. Ciach, O. Patsahan]{A. Ciach\refaddr{label1},
       O. Patsahan\refaddr{label2}}
\addresses{
\addr{label1} Institute of Physical Chemistry,
 Polish Academy of Sciences, 01-224 Warszawa, Poland
\addr{label2} Institute for Condensed Matter Physics of the National
Academy of Sciences of Ukraine, \\ 1 Svientsitskii Str., 79011 Lviv, Ukraine }

\authorcopyright{A. Ciach, O. Patsahan, 2012}
\date{Received December 31, 2011, in final form March 6, 2012}

\begin{document}

\maketitle

\begin{abstract}
Equation of state for systems with particles self-assembling into
aggregates is derived within a mesoscopic theory
 combining density functional and field-theoretic approaches. We focus on the effect of mesoscopic fluctuations in the
 disordered phase. The pressure -- volume fraction isotherms are calculated explicitly
for two forms of the short-range attraction long-range repulsion
potential. Mesoscopic fluctuations lead to an increased pressure in
each case, except for very small volume fractions. When large
clusters are formed, the mechanical instability of the system is
present at much higher temperature than found in mean-field
approximation. In this case phase separation competes with the formation
of periodic phases (colloidal crystals). In the case of small clusters, no
mechanical instability associated with separation into dilute and
dense phases appears.
\pacs 61.20.Gy, 64.10.+h, 64.60.De, 64.75.Yz
\end{abstract}

\section{Introduction}
Recent
experimental~\cite{stradner:04:0,porcar:00:0,campbell:05:0,stiakakis:05:0},
theoretical~\cite{sear:99:0,archer:07:0,archer:08:0,ciach:08:1,ciach:10:1,ciach:11:0,archer:07:1}
and
simulation~\cite{sciortino:04:0,sciortino:05:0,archer:07:1,candia:06:0,toledano:09:0,kowalczyk:11:0}
studies reveal that in many systems with competing interactions,
clusters or aggregates of different sizes and shapes are formed. These objects in certain thermodynamic states can form ordered structures in
space~\cite{ciach:08:1,ciach:10:1,candia:06:0}. Notable examples
include charged globular proteins in
water~\cite{stradner:04:0,shukla:08:0,porcar:00:0,kowalczyk:11:0},
and mixtures of small nonadsorbing polymers with charged colloids or
micelles~\cite{stradner:04:0,campbell:05:0,zhang:09:0}. Interactions
(or in fact effective interactions) in the latter systems can be
described by the model potential consisting of short-range
attraction, resulting from solvophobic or depletion interactions,
and long-range repulsion, resulting from screened electrostatic
potential (SALR potential).

Systems containing clusters or aggregates are inhomogeneous on the length scale associated with the average size of
the aggregates and average distance between them. The corresponding length scale of the inhomogeneities is significantly
 larger than the size of the particles. Fluctuations on the mesoscopic length scale corresponding to displacements of
the aggregates have an important impact on the grand potential, and thus on the equation of state (EOS). Derivation of
 an accurate EOS for inhomogeneous systems is less trivial than in the case of homogeneous systems, since it is
necessary to perform summation over different spatial distributions of the clusters and over all deformations of them.

Contribution to the grand potential associated with mesoscopic fluctuations can be calculated in the field-theoretic
approach~\cite{ciach:08:1,ciach:10:1}. In principle, this contribution can be obtained in the perturbation expansion in
terms of Feynman diagrams. In practice, an approximate result can be analytically obtained in the self-consistent Hartree
approximation~\cite{ciach:06:0,ciach:08:1,ciach:11:0}.
A formal expression for the fluctuation contribution to the grand
potential has been derived in
references~\cite{ciach:06:0,ciach:08:1,ciach:11:0}.
However, its explicit form with the chemical potential expressed in terms of temperature and density  has not been  determined yet. The EOS isotherms for various forms of the SALR potential were not analyzed, and the effect of the mesoscopic
 fluctuations on pressure remains an open question.

\begin{figure}[ht]
\centering
\includegraphics[scale=0.35]{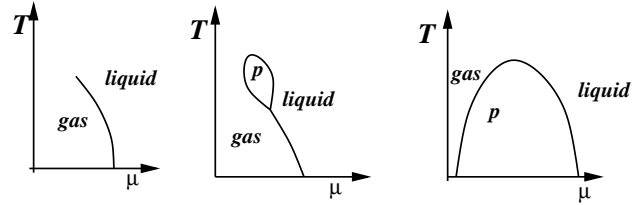}
\caption{ Types of possible phase diagrams for the SALR potentials, shown schematically.
From the left to the right panel the role of the repulsion increases.
`p' indicates the stability region of the periodic phases of different symmetries.
 The disordered fluid close to the stability region of the periodic phases is inhomogeneous, i.e. clusters are formed,
 but their positions are correlated only at short distances. In this work we are interested in systems exhibiting
 the phase behaviour shown in the right (System 1) and the central (System 2) panel.                     }\label{fig1}
\end{figure}

It is important to note that
 various forms of the SALR potential are associated with different properties of the systems. Depending on the ratios
 between the strengths and ranges of the attractive and repulsive parts of the potential, separation into uniform phases,
 formation of clusters of various sizes and shapes (globules, cylinders, slabs) in the so-called microsegregation, or isolated individual particles may occur. In certain conditions, the clusters can be periodically distributed in space in the periodic phases whose densities are smaller than the density of the liquid phase~\cite{candia:06:0,ciach:08:1,ciach:10:1}.
 Possible types of the phase diagram for different SALR potentials are shown in figure~\ref{fig1} (see also references~\cite{archer:07:0,archer:08:0,ciach:08:1}).
Properties of the disordered phase can be influenced by the  periodic phases for thermodynamic states close to the
stability of the latter. We expect that the disordered phase, although the long-range order is absent,
is inhomogeneous on the mesoscopic length scale and resembles `molten periodic phases'. In this respect the inhomogeneous  disordered phase is similar to microemulsion which can be interpreted as molten  lyotropic liquid crystal.

In this work we focus on the stable or metastable disordered inhomogeneous phase in which clusters are formed, but they do not form any ordered structure. We derive the EOS with the contribution from mesoscopic length-scale fluctuations included. We calculate the explicit form of the EOS for two
representative examples of the SALR potential within the self-consistent Hartree approximation. The first system
 corresponds to the formation of small clusters, and the gas-liquid separation
 is unstable for all temperatures in the mean-field (MF) approximation (figure~\ref{fig1}, right panel).
 In the second system, large clusters are formed. The gas-liquid separation is
present in this system as a stable or a metastable
transition for  low temperatures (figure~\ref{fig1}, central panel). We shall compare the effect of mesoscopic
fluctuations in these two cases on the isotherms $P(\zeta)$, where $\zeta$ is the volume fraction of particles and $P$
is pressure.

In the next section we briefly summarize the mesoscopic approach. In section~3 the EOS is obtained by two methods.
In section~3.1 we consider mesoscopic fluctuations
about the average volume fraction, while in section~3.2 fluctuations about the most probable volume fraction are included.
The formulas derived in section~3 are
evaluated for the two versions of the SALR potential in section~4.   We obtain a completely different effect of the mesoscopic fluctuations in these two cases.
The two approaches (sections~3.1 and 3.2)  yield very close results provided that the relative fluctuation contribution to the average volume fraction is small.
For larger fluctuation-induced shifts of the volume fraction,  only qualitative agreement of the two methods is obtained. Short summary is presented in section~5.

\section{Short summary of the mesoscopic description}

We consider a local volume fraction of particles, i.e. the microscopic volume fraction averaged over mesoscopic regions, as an order parameter~\cite{ciach:11:0}.
The corresponding
mesoscopic volume fraction varies on a length scale larger than the size of the particles, and the
characteristic size of inhomogeneities is the upper limit for the mesoscopic length scale. A particular form of the mesoscopic volume fraction  can be considered as a constraint on the microscopic states. The corresponding mesostate
is a subset of microstates compatible with the imposed constraint. The mesoscopic volume fraction (or the mesostate)
was defined  in references~\cite{ciach:08:1,ciach:11:0}. For a one-component case
we fix the mesoscopic length scale $R\geqslant \sigma/2$ and consider spheres $S_R({\mathbf{r}})$ of radius $R$ and centers at $\mathbf{r}$ that cover the whole
volume $V$ of the system. We define the {\it mesoscopic} volume fraction  at $\mathbf{r}$ by
\begin{equation}
\label{mesor}
 \zeta(\mathbf{r}):=\frac{1}{V_S} \int_{\mathbf{r'}\in S_R(\mathbf{r})}\hat\zeta(\mathbf{r'},{\cal M}),
\end{equation}
where $V_S=4\pi R^3/3$, and the  {\it microscopic} volume fraction in the microstate  ${\cal M}= \{\{{\bf r}_i\}_{i=1,\ldots,N}\}$ is
 defined by
\begin{equation}
\label{microe}
 \hat\zeta({\bf r},{\cal M}):=
\sum_{i=1}^{N}\theta\left(\frac{\sigma}{2}-|{\bf r}-{\bf r}_i|\right),
\end{equation}
where
 $\theta(r)$ is the Heaviside unit step function. The microscopic volume fraction is equal to $1$ at points that are inside one of the hard spheres, and zero otherwise. Integrated over the system volume, it yields the volume occupied by the particles. The mesoscopic volume fraction at ${\bf r}$ is equal to the fraction of the volume of the sphere  $S_R({\bf r})$  that is occupied by the particles. Note that $\zeta({\bf r})$ takes the same value if one particle is entirely included in this sphere, independently of the precise position of its centre. Thus, $\zeta({\bf r})$ gives less precise information on the distribution of particles than $\hat\zeta({\bf r})$. In the disordered phase  $\zeta({\bf r})$ is independent of
${\bf r}$ and equals the fraction of the total volume that is occupied by the particles.
 The mesostate can be imagined as a fixed distribution of centers of
 clusters, with arbitrary distribution of particles within the clusters, and  small modifications of their shapes.
 Probability of the mesostate $\zeta$ is given by~\cite{ciach:08:1,ciach:11:0}
 \begin{equation}
\label{p2}
  p[\zeta]=\frac{\re^{-\beta\Omega_\mathrm{co}[\zeta]}}{\Xi}\,,
 \end{equation}
where
\begin{equation}
\label{Xi1}
 \Xi=\int\nolimits' D\zeta \, \re^{-\beta\Omega_\mathrm{co}[\zeta]}.
\end{equation}
The functional integral $ \int^{'}D\zeta $ in (\ref{Xi1}) is over
all mesostates,
\begin{equation}
\label{Omco}
\Omega_\mathrm{co}=U[\zeta]-TS[\zeta]-\mu N[\zeta],
\end{equation}
 where $U,S, N$ are the internal energy, entropy and the number of molecules respectively in the system with the constraint
 of compatibility with the mesostate $[\zeta]$ imposed on the microscopic volume fractions.  $U$ is given by the well known
 expression
\begin{equation}
\label{U}
 U[\zeta]=\frac{1}{2}\int_{\bf r_1}\int_{\bf r_2}V_\mathrm{co}({\bf r}_1-{\bf r}_2)\zeta({\bf r}_1)\zeta({\bf r}_2),
\end{equation}
where for spherically symmetric interactions
\begin{eqnarray}
\label{Vco}
 V_\mathrm{co}({\bf r}_1-{\bf r}_2)= V(r_{12})g_\mathrm{co}({\bf r}_1-{\bf r}_2),
\end{eqnarray}
 $r_{12}=|{\bf r}_1-{\bf r}_2|$, $v^2 V(r_{12})$ is the interaction potential, $v=\pi\sigma^3/6$ is the volume of the particle,
 and $g_\mathrm{co}({\bf r}_1-{\bf r}_2)$   is the microscopic pair correlation function
for the microscopic volume fraction in the system with the constraint of compatibility with the mesostate imposed
 on the microscopic states.
The grand potential can be written in the form
\begin{equation}
\label{FF}
 \beta \Omega[\bar\zeta]=\beta\Omega_\mathrm{co}[\bar\zeta]-\log
 \Xi_\mathrm{fluc}\,,
\end{equation}
where
\begin{equation}
\Xi_\mathrm{fluc}=\int D\phi\, \re^{-\beta H_\mathrm{fluc}[\bar\zeta,\phi]},
\end{equation}
\begin{eqnarray}
\label{Hfl}
H_\mathrm{fluc}[\bar\zeta,\phi]=\Omega_\mathrm{co}[\bar\zeta+\phi]-\Omega_\mathrm{co}[\bar\zeta]=
\sum_{n=1}\int_{\bf r_1}\ldots\int_{\bf r_n}\frac{C^\mathrm{co}_n[\bar\zeta]}{n!}\phi({\bf r}_1)\ldots\phi({\bf r}_n).
\end{eqnarray}
The average mesoscopic volume fraction, $\bar\zeta$, corresponds to  the minimum of $\Omega$, and must satisfy the equation
\begin{eqnarray}
\label{avdg}
 \frac{\delta\beta\Omega_\mathrm{co}[\bar\zeta]}{\delta\bar\zeta({\bf r}) }+
\left\langle \frac{\delta(\beta H_\mathrm{fluc})}{\delta\bar\zeta({\bf r})}\right\rangle=0,
\end{eqnarray}
where the averaging is over the fields $\phi$ with the probability $\propto\exp(-\beta H_\mathrm{fluc}[\bar\zeta,\phi])$.
Note that when \linebreak $C^\mathrm{co}_n[\bar\zeta]=0$ for odd $n$, then the second term on the LHS in (\ref{avdg}) vanishes, and the average volume fraction
coincides with the most probable volume fraction $\zeta_0$ given by
\begin{eqnarray}
\label{avdg0}
 \left.\frac{\delta\beta\Omega_\mathrm{co}[\zeta]}{\delta\zeta({\bf r}) }\right\vert_{\zeta=\zeta_0}=0.
\end{eqnarray}
By contrast, when $C^\mathrm{co}_n[\bar\zeta]\ne 0$ for odd $n$, then $\bar\zeta\ne \zeta_0$.

In order to evaluate the fluctuation contribution to $\Omega[\bar\zeta]$
 we decompose $H_\mathrm{fluc}[\bar\zeta,\phi]$ into two parts
\begin{equation}
\label{Har}
H_\mathrm{fluc}[\bar\zeta,\phi] ={\cal
H}_\mathrm{G}[\bar\zeta,\phi]+\Delta{\cal H}[\bar\zeta,\phi],
\end{equation}
where in the disordered phase
\begin{eqnarray}
\label{Haga}
 {\cal H}_\mathrm{G}[\bar\zeta,\phi]=
\frac{1}{2}\int_{\bf k}\tilde \phi({\bf k})\tilde C_2({\bf k},\bar\zeta)\tilde\phi(-{\bf k}),
\end{eqnarray}
and $\tilde C_2({\bf k},\zeta)$ is the Fourier transform of
\begin{equation}
\label{calC2}
 C_2(r_{12},\zeta)= \frac{\delta^2 \beta \Omega[\zeta]}{\delta\zeta({\bf r}_1)\delta\zeta({\bf r}_2)}\,.
 \end{equation}
The above function calculated for  $\zeta=\bar\zeta$ is related to the direct correlation function~\cite{evans:79:0}.

Assuming $\Delta{\cal H}\ll {\cal H}_\mathrm{G}$,
we obtain \cite{ciach:06:0,ciach:08:1}
\begin{eqnarray}
\label{OmHarval} \beta\Omega[\bar\zeta]\approx
\beta\Omega_\mathrm{co}[\bar\zeta]-\log\int D\phi \, \re^{-\beta{\cal H}_\mathrm{G}}
+\langle \beta\Delta{\cal H}\rangle_\mathrm{G} +O\left(\langle \beta\Delta{\cal
H}\rangle_\mathrm{G}^2\right),
\end{eqnarray}
where $\langle \ldots\rangle_\mathrm{G}$ denotes the averaging with the Gaussian Boltzmann
factor $\propto \re^{-\beta{\cal H}_\mathrm{G}}$.

Since $P=-\Omega/V$, approximate EOS can be obtained from (\ref{OmHarval}) calculated for $\bar\zeta$ satisfying (\ref{avdg}),
 when $\Omega_\mathrm{co}[\bar\zeta]$ (see (\ref{Omco}))  is known.
 The chemical potential in (\ref{OmHarval}) should be expressed in terms of $T$ and $\bar\zeta$; its form  as a function of
 $T$ and $\bar\zeta$ can be  determined from equation~(\ref{avdg}). In order to evaluate the second term in equation~(\ref{avdg}), we
 need approximate forms of the correlation functions. In the lowest order approximation, it is necessary to  determine
\begin{equation}
\label{GC} \langle\tilde\phi({\bf k})\tilde\phi(-{\bf
k})\rangle=\tilde G_2(k,\bar\zeta)=1/\tilde C_2(k,\bar\zeta).
 \end{equation}

\section{ Approximate results for the fluctuation contributions to the EOS, density and chemical potential}

\looseness=-1In this section we derive an explicit form of the EOS $P(\bar\zeta,T)$, the density shift,
$\Delta\zeta=\bar\zeta-\zeta_0$, and the chemical potential $\mu(\bar\zeta,T)$
under the following assumptions:
 (i) local density approximation for $S[\zeta]$
and (ii) the lowest-order approximation for the second term in (\ref{avdg}). In the local density approximation we have
\begin{eqnarray}
\label{Fhfh}
 -TS[\zeta]=\int_{\bf r}f_\mathrm{h}(\zeta({\bf r})),
\end{eqnarray}
where $f_\mathrm{h}(\zeta)$ is the free-energy density of the hard-sphere system with dimensionless density $\rho^*= 6\zeta/\pi$.
We  assume the Percus-Yevick  approximation
\begin{eqnarray}
\label{PY}
\beta f_\mathrm{h}(\zeta)=\rho^*\ln(\rho^*)-\rho^*+
\rho^*\Bigg[\frac{3\zeta(2-\zeta)}{2(1-\zeta)^2}-\ln(1-\zeta)\Bigg].
\end{eqnarray}
In the local density approximation  $C_n^\mathrm{co}[\zeta]$ are just functions of $\zeta$  in the disordered phase,
and we can simplify the notation, introducing
\begin{eqnarray}
A_n(\zeta)= \frac{\rd^n [\beta f_\mathrm{h} (\zeta)]}{\rd \zeta^n}\,.
\end{eqnarray}
 For $n>2$ we have
\begin{eqnarray}
\label{local}
 C_n^\mathrm{co}[\zeta]=A_n(\zeta),
\end{eqnarray}
  whereas for $n=2$
\begin{eqnarray}
\label{Cco}
 \tilde C^\mathrm{co}_2(k,\zeta)= \beta\tilde V_\mathrm{co}(k)+A_2(\zeta),
\end{eqnarray}
where in the disordered phase $\tilde C_2^\mathrm{co}(k,\zeta)$ is the Fourier transform of the function
\begin{eqnarray}
\label{Co}
 C^\mathrm{co}_2(r_{12},\zeta)= \frac{\delta^2 \beta \Omega_\mathrm{co}[\zeta]}{\delta\zeta({\bf r}_1)\delta\zeta({\bf r}_2)}\,.
\end{eqnarray}
Equations~(\ref{PY}) and (\ref{U}) define the functional $\Omega_\mathrm{co}$ for a given form of  $V_\mathrm{co}$ (equation~(\ref{Vco})).

From equation~(\ref{p2}) it follows that the most probable fluctuations correspond to the wavenumbers $k=k_\mathrm{b}$ for which
 $\tilde V_\mathrm{co}(k)$ assumes the minimum, and the inhomogeneities on the length scale $2\pi/k_\mathrm{b}$ are energetically
favored when $\tilde V_\mathrm{co}(k_\mathrm{b})<0$. In this work we focus on the
effect of the self-assembly into aggregates. Therefore, we restrict
our attention to
 $\tilde V_\mathrm{co}(k)$ which assumes the minimum for $k_\mathrm{b}>0$, and  $\tilde V_\mathrm{co}(k_\mathrm{b})<0$~\cite{ciach:08:1,ciach:10:1}.
 Since the fluctuations with the wavenumber $k\approx k_\mathrm{b}$ are most probable, they yield the main
fluctuation contribution to the grand potential~(\ref{OmHarval}). For such fluctuations we can make the approximation
\begin{eqnarray}
\label{VBV}
\beta \tilde V_\mathrm{co}(k)\approx \beta \tilde V_\mathrm{co}(k_\mathrm{b})+\beta \tilde V_\mathrm{co}^{(2)}(k_\mathrm{b})(k-k_\mathrm{b})^2/2+\ldots
\end{eqnarray}
As the energy scale we choose the excess energy associated with the fluctuations having unit amplitude and the wavenumber
$k_\mathrm{b}$, and introduce the notation
 \begin{eqnarray}
\label{T*}
  \beta^*=1/T^*= \beta |\tilde V_\mathrm{co}(k_\mathrm{b})|,
 \end{eqnarray}
\begin{eqnarray}
\label{v2*}
 v_2^*=\frac{\tilde V_\mathrm{co}^{(2)}(k_\mathrm{b})}{2|\tilde V_\mathrm{co}(k_\mathrm{b})|}
\end{eqnarray}
and
\begin{eqnarray}
\label{v0*}
 v_0^*=\frac{\tilde V_\mathrm{co}(0)}{\tilde V_\mathrm{co}(k_\mathrm{b})}\,.
\end{eqnarray}

\subsection{Fluctuations around the average volume fraction}

In this subsection we consider fluctuations about the average value $\bar\zeta$. We shall first determine the chemical potential as a function of $\bar\zeta$ and $T^*$ from (\ref{avdg}). In order to calculate the second term in (\ref{avdg}),
we assume that relevant fluctuations are of small amplitudes, and truncate the expansion in (\ref{Hfl})
at the fourth-order term. Next we insert the derivative with respect to $\bar\zeta$ of the RHS of equation~(\ref{Hfl})
 truncated at the quadratic term in $\phi$, and we obtain from (\ref{avdg}) and (\ref{Omco})  an approximate equation for
 the rescaled chemical potential, $\bar\mu=6 \mu/(\pi\sigma^3)$, of the form
\begin{eqnarray}
\label{barmu}
 \beta \bar\mu\approx \beta\bar\mu^\mathrm{MF}(\bar\zeta)+\frac{A_3 (\bar\zeta)}{2}{\cal
 G}(\bar\zeta),
\end{eqnarray}
where
\begin{eqnarray}
\label{barmuMF}
  \beta\bar\mu^\mathrm{MF}(\bar\zeta)=\beta\tilde V_\mathrm{co}(0) \bar\zeta +\beta f_\mathrm{h}'(\bar\zeta)
\end{eqnarray}
and the  last term in (\ref{barmu})  is the fluctuation contribution with
\begin{eqnarray}
\label{calG}
{\cal G}(\zeta)=\int_{\bf k}\tilde G_2(k,\zeta)=G_2(0,\zeta).
\end{eqnarray}
The same expression can be obtained from $\langle \phi\rangle=0$, when (\ref{Har}), (\ref{Haga}) and the approximation
\begin{eqnarray}
\label{aH}
  \re^{-\beta H_\mathrm{fluc}} =\re^{-\beta H_\mathrm{G}}\Big[1-\beta \Delta H +O\left(\Delta H^2\right)\Big]
\end{eqnarray}
 are used.
The approximation (\ref{barmu}) is valid as long as the correction
term is not larger than the MF result. When considering particular
cases we shall verify if this is the case.
 The fluctuation contribution to the direct correlation function~(\ref{calC2})
 is obtained by calculating the second derivative of the second term on the RHS of~(\ref{FF}) with respect to $\bar\zeta$.
In the consistent approximation
we insert in the obtained expression the appropriate derivatives of $H_\mathrm{fluc}$ (equation~(\ref{Hfl}))
with the expansion in $\phi$ truncated at the second order. The result is
given by~\cite{ciach:06:0,patsahan:07:0,ciach:08:1,ciach:11:0}
\begin{eqnarray}
\label{BrazC}
 \tilde C_2(k,\zeta)\approx\tilde C^\mathrm{co}_2(k,\zeta)+\frac{A_4(\zeta)}{2}{\cal G}(\zeta).
\end{eqnarray}
Equations~(\ref{BrazC}), (\ref{GC})  and (\ref{calG}) should be solved self-consistently.

 The fluctuation induced shift of the volume fraction,
$\Delta \zeta=\bar\zeta-\zeta_0$, can be obtained from~(\ref{avdg}) by expanding the first term on the LHS
about $\zeta_0$,
 \begin{eqnarray}
\label{shift} \tilde
C_2^\mathrm{co}(0,\zeta_0)\Delta\zeta+\sum_{n=2}^{\infty}\frac{
C_{n+1}^\mathrm{co}[\zeta_0]}{n!}\Delta \zeta^{n}=- \left\langle
\frac{\delta(\beta H_\mathrm{fluc})}{\delta\bar\zeta({\bf r})}\right\rangle.
\end{eqnarray}
 For small  $\Delta \zeta$ we can truncate the expansion in~(\ref{shift}) at the first term.
When the RHS in~(\ref{shift}) is approximated  as in the calculation of $\bar\mu$ from~(\ref{avdg}), we obtain the result
\begin{eqnarray}
\label{Dzeta}
 \Delta\zeta\approx -\frac{A_3(\zeta_0)}{2\tilde C_2(0,\zeta_0)}{\cal G}(\bar\zeta)
\approx -\frac{A_3(\zeta_0)}{2\tilde C_2(0,\zeta_0)+A_3(\zeta_0){\cal G}'(\zeta_0)}{\cal G}(\zeta_0).
\end{eqnarray}
We used the approximations:
 $A_3(\bar\zeta)=A_3(\zeta_0)+A_4(\zeta_0)\Delta\zeta +O\left(\Delta\zeta^2\right)$,
 ${\cal G}(\bar\zeta)={\cal G}(\zeta_0)+{\cal G}'(\zeta)\Delta\zeta +O\left(\Delta\zeta^2\right)$ and equation~(\ref{BrazC}).

For the potential given in
 (\ref{VBV}) the approximate form of ${\cal G}$ is \cite{brazovskii:75:0,ciach:06:0,patsahan:07:0,ciach:08:1}
\begin{eqnarray}
\label{calG3}
{\cal G}(\zeta)=\frac{2a\sqrt{T^*}}{Z(\zeta)}\,,
\end{eqnarray}
where
\begin{equation}
\label{a}
a=\frac{k_\mathrm{b}^2}{4\pi\sqrt {v_2^*}}
\end{equation}
and
$Z(\zeta)=\sqrt{\tilde C_2(k_\mathrm{b},\zeta)}$. The above approximation is valid for
 $\tilde C_2(k_\mathrm{b},\zeta)\ll\beta^*v_2^*k_\mathrm{b}^2$ \cite{brazovskii:75:0,ciach:06:0,patsahan:07:0,ciach:08:1}.
The equation (\ref{BrazC}) for $k=k_\mathrm{b}$  takes the form
\begin{equation}
\label{r}
 Z(\zeta)^3=Z(\zeta)\tilde C_2^\mathrm{co}(k_\mathrm{b},\zeta)+A_4(\zeta) a\sqrt{T^*},
\end{equation}
 and the explicit expression for $Z$ is
\begin{equation}
 Z(\zeta)=\frac{W(\zeta)}{6}+\frac{2\tilde C_2^\mathrm{co}(k_\mathrm{b},\zeta)}{W(\zeta)}
\end{equation}
with
\begin{equation}
 W(\zeta)=\left\{
108 A_4(\zeta) a\sqrt{T^*} +12\sqrt{-12 \tilde C_2^\mathrm{co}(k_\mathrm{b},\zeta)^3 +81\left[A_4(\zeta) a\sqrt{T^*}\right]^2}
\right\}^{1/3}.
\end{equation}

 The fluctuation contribution
in equation~(\ref{OmHarval}) for the approximations (\ref{BrazC})--(\ref{r}) was calculated in references~\cite{brazovskii:75:0,ciach:06:0,patsahan:07:0,ciach:08:1}, and has the form
\begin{equation}
 \label{OmHarval1}
\beta\Omega[\bar\zeta]\approx \beta\Omega_\mathrm{co}[\bar\zeta]+2a\sqrt{T^*}Z(\bar\zeta)V-
\frac{A_4(\bar\zeta){\cal G}(\bar\zeta)^2}{8}V.
\end{equation}
Taking into account (\ref{barmu}) and (\ref{PY}), we obtain
from  (\ref{OmHarval1})  the explicit form of the EOS
\begin{equation}
\label{EOS}
\beta P(\bar\zeta)=\beta P^\mathrm{MF}(\bar\zeta)
+
F(\bar\zeta, T^*),
\end{equation}
where
\begin{equation}
\label{EOSMF}
\beta P^\mathrm{MF}(\bar\zeta)=-\frac{\beta^* v_0^*}{2}\bar\zeta^2+\bar\zeta\frac{\rd\beta f_\mathrm{h}(\bar\zeta)}{\rd\bar\zeta}-\beta f_\mathrm{h}(\bar\zeta)=
-\frac{\beta^* v_0^*}{2}\bar\zeta^2 +\rho^*\frac{\bar\zeta^2+
\bar\zeta+1}{\left(1-\bar\zeta\right)^3}
\end{equation}
and
\begin{eqnarray}
\label{F}
 F(\bar\zeta, T^*)&=&\frac{a\sqrt{T^*} A_3(\bar\zeta)\bar\zeta}{Z(\bar\zeta)}-2a\sqrt{T^*} Z(\bar\zeta)+
\frac{a^2 A_4(\bar\zeta) T^*}{2 Z(\bar\zeta)^2}  \nonumber \\
&=&
\frac{a\sqrt{T^*}}{2}\left[
\frac{2A_3(\bar\zeta)\bar\zeta -\tilde C_2^\mathrm{co}(k_\mathrm{b},\bar\zeta)}{Z(\bar\zeta)}-3Z(\bar\zeta)
\right].
\end{eqnarray}
The second equality in (\ref{EOSMF}) is valid for the PY approximation for $f_\mathrm{h}$.
In order to obtain the last equality in (\ref{F}), equation~(\ref{r}) was used.

\subsection{Fluctuations around the most probable volume fraction}

In the previous subsection we considered fluctuations about the average value, which in general differs from the most probable value of the volume fraction.
In principle, it is possible to consider equations analogous to~(\ref{FF}) and~(\ref{Hfl}), but with $\bar\zeta$
replaced by $\zeta_0$. In this new approach equation~(\ref{avdg0}) is satisfied, and thus the expansion in (\ref{Hfl})
starts with $n=2$ ($C_1^\mathrm{co}[\zeta_0]=0$). On the other hand, $\langle\phi\rangle=\bar\zeta-\zeta_0\ne 0$. The results
obtained in the two approaches~--- with included fluctuations around the average value or around the most probable value~---
should be the same in the exact theory. However,   when the fluctuation contribution is obtained in an approximate theory,
the results may depend on the validity of the assumptions made in the two approaches. In this section we derive an alternative
version of the EOS, based on the contribution from the fluctuations around the most probable value.
From (\ref{avdg0}) we obtain for the chemical potential
\begin{eqnarray}
\label{barmu0}
 \bar\mu(\zeta_0)= \bar\mu^\mathrm{MF}(\zeta_0)
\end{eqnarray}
with $\bar\mu^\mathrm{MF} $ given in equation~(\ref{barmuMF}).
 We consider (\ref{Har}) with $\bar\zeta$ replaced by $\zeta_0$, and the approximation (\ref{aH}).
In the above, $  H_\mathrm{G}$ is given in equation~(\ref{Haga}) with $\tilde C_2(k,\zeta_0)=1/\tilde G_2(k,\zeta_0)$, where $\tilde G_2(k,\zeta_0)=\langle\tilde \phi({\bf k})\tilde \phi(-{\bf k})\rangle$, and
 \begin{eqnarray}
\label{dH}
\Delta H&=&\frac{1}{2}\int_{\bf r'} \int_{\bf r''} \phi({\bf r}')\left[C_2^\mathrm{co}\left({\bf r}'-{\bf r}'',\zeta_0\right)-C_2\left({\bf r}'-{\bf r}'',\zeta_0\right)\right] \phi({\bf r}'')%
+\int_{\bf r}\frac{A_3(\zeta_0)}{3!}\phi({\bf r})^3
+\int_{\bf r}\frac{A_4(\zeta_0)}{4!}\phi({\bf r})^4 +\ldots\nonumber\\
\end{eqnarray}
 Note that the function $\tilde G_2(k,\zeta_0)$  defined here differs from the  correlation function,
because in this case $\langle\phi({\bf r})\rangle\ne 0$. Taking into account that for $H_\mathrm{G}$ of the form (\ref{Har}) there holds $\int D\phi \re^{-\beta H_\mathrm{G}}\phi^{2n+1}=0$,  we obtain the expression
\begin{eqnarray}
\label{phi}
 \langle\phi({\bf r})\rangle\approx \frac{-\int D\phi \,
\frac{1}{3!}\int_{\bf r'}\int_{\bf r''}\int_{\bf r'''}A_3(\zeta_0)
\phi\left({\bf r}'\right)\phi\left({\bf r}''\right)\phi\left({\bf r}'''\right)\phi({\bf r})\re^{-\beta H_\mathrm{G}}}{\Xi_\mathrm{fluc}}\,.
\end{eqnarray}
Finally,  the lowest-order result is
\begin{eqnarray}
\label{shift2}
 \bar\zeta=\zeta_0+\langle\phi\rangle \approx \zeta_0-\frac{A_3(\zeta_0)}{2\tilde C_{2}(0,\zeta_0)}{\cal G}(\zeta_0).
\end{eqnarray}
At the same level of
approximation $\tilde C_2(k,\zeta_0)$  is given in equation~(\ref{BrazC}),
 except that all quantities are calculated at $\zeta_0$ which satisfies~(\ref{avdg0}) rather than~(\ref{avdg}).
This can be verified by a direct calculation of $\langle\phi({\bf r}_1) \phi({\bf r}_2)\rangle$ with the help of~(\ref{aH}) and~(\ref{dH}), in an  approximation
 analogous to (\ref{phi}) (see reference~\cite{patsahan:07:0}).

The above shift of the volume fraction differs from~(\ref{Dzeta}), because instead of
 ${\cal G}(\bar\zeta)$, there appears ${\cal G}(\zeta_0)$.
The dependence of $\bar\mu$ on the average volume fraction is given in equations~(\ref{barmu0}) and (\ref{shift2}), with eliminated~$\zeta_0$.

In order to evaluate the EOS, we consider an equation analogous to (\ref{OmHarval1}),
with $\Omega_\mathrm{co}[\zeta_0]$ calculated at its minimum $\zeta=\zeta_0$.
The EOS takes the form
\begin{equation}
\label{EOS0}
\beta P(\zeta_0)=\beta P^\mathrm{MF}(\zeta_0) +F_0(\zeta_0,
T^*),
\end{equation}
where $\beta P^\mathrm{MF}$ is defined in (\ref{EOSMF}), $\zeta_0 $ satisfies (\ref{avdg0}), and
\begin{eqnarray}
\label{F0}
 F_0(\zeta_0, T^*)=-2a\sqrt {T^*} Z(\zeta_0)+\frac{a^2 A_4 (\zeta_0)T^*}{2
 Z(\zeta_0)^2}\,.
\end{eqnarray}
The dependence of $P$ on the average volume fraction $\bar\zeta$ is given by parametric equations (\ref{EOS0}) with (\ref{F0})
 and (\ref{shift2}).

The approximate theory developed in this section is valid for small $\Delta\zeta$ since we assumed that the relevant fluctuations are small and truncated the expansion in (\ref{Hfl}) at the term $\propto \phi^4$. Moreover, to evaluate $\Delta\zeta$ we neglected the terms of the order $O\left(\Delta \zeta^2\right)$. We may expect that if we obtain large $\Delta\zeta$ and large discrepancies between the results obtained by the two methods, then the approximate theory is not sufficiently accurate.

\subsection{Comparison between the two methods}

 Let us  focus on the chemical potential, and compare the two expressions, equations~(\ref{barmu}) and (\ref{barmu0})
where $\bar\zeta$ and $\zeta_0$ satisfy equations~(\ref{avdg}) and (\ref{avdg0}), respectively. We expand the RHS in equation~(\ref{barmu}) about $\zeta_0$,
\begin{eqnarray}
\bar\mu&\approx& \beta\tilde V_\mathrm{co}(0) \zeta_0 +\beta f_\mathrm{h}'(\zeta_0)
+ \Delta\zeta \Bigg[\beta \tilde V_\mathrm{co}(0) +A_2(\zeta_0)+ \frac{A_4(\zeta_0)}{2}{\cal G}(\zeta_0)
\nonumber \\
&&+\frac{A_3(\zeta_0)}{2}{\cal G}'(\zeta_0)\Bigg]
+\frac{A_3(\zeta_0)}{2}{\cal G}(\zeta_0) +O\left( \Delta\zeta^2\right).
\end{eqnarray}
From (\ref{Dzeta}), (\ref{BrazC}) and (\ref{barmu0}) we obtain an equality of the two expressions for the chemical potential
 to the linear order in $ \Delta\zeta$, when  $ \Delta\zeta$ is given in  (\ref{Dzeta}).

Similarly, to compare the two expressions for the EOS, equations~(\ref{EOS}) and (\ref{EOS0}), we expand the RHS of equation~(\ref{EOS}) about $\zeta_0$ to the linear
order in $\Delta \zeta$.
Taking into account~(\ref{Dzeta}) and~(\ref{r}), we arrive at equation~(\ref{EOS0}), up to the terms proportional to $A_5$. The
 latter are disregarded in an approximation consistent with the $\phi^4$ theory for the fluctuation contribution considered
 in this work.  For relatively large
 $\Delta \zeta$, when the terms beyond the linear order become important, discrepancies between the results obtained by the
two methods should be expected.

\section{Explicit results for two model potentials}

In this subsection we shall compare the expressions for the chemical potential and for the pressure obtained by the two approaches for two systems showing a qualitatively different behavior.
We shall evaluate the EOS~(\ref{EOS}) for the representative model SALR potential,
\begin{equation}
\label{int_pot_r} V_\mathrm{co}(r)=\left[-\frac{{\cal
A}_1}{r}\re^{-z_1r}+\frac{{\cal A}_2}{r}\re^{-z_2r}\right]\theta(r-1),
\end{equation}
where $z_i$ is the inverse range in
$\sigma^{-1}$ units. The function $\theta(r-1)$ is a very crude approximation for the pair distribution function.
In  Fourier representation, the above SALR potential takes the form
\begin{equation}
\label{uLRIIF}
\tilde V_\mathrm{co}(k)=4\pi\left[\frac{{\cal A}_2\re^{-z_2}}{z_2^2+k^2}
\left(z_2\frac{\sin k}{k}+\cos k\right)-\frac{{\cal A}_1\re^{-z_1}}{z_1^2+k^2}
\left(z_1\frac{\sin k}{k}+\cos k\right)\right].
\end{equation}

\begin{figure}[ht]
\centerline{
\hspace{5mm}\includegraphics[scale=0.32]{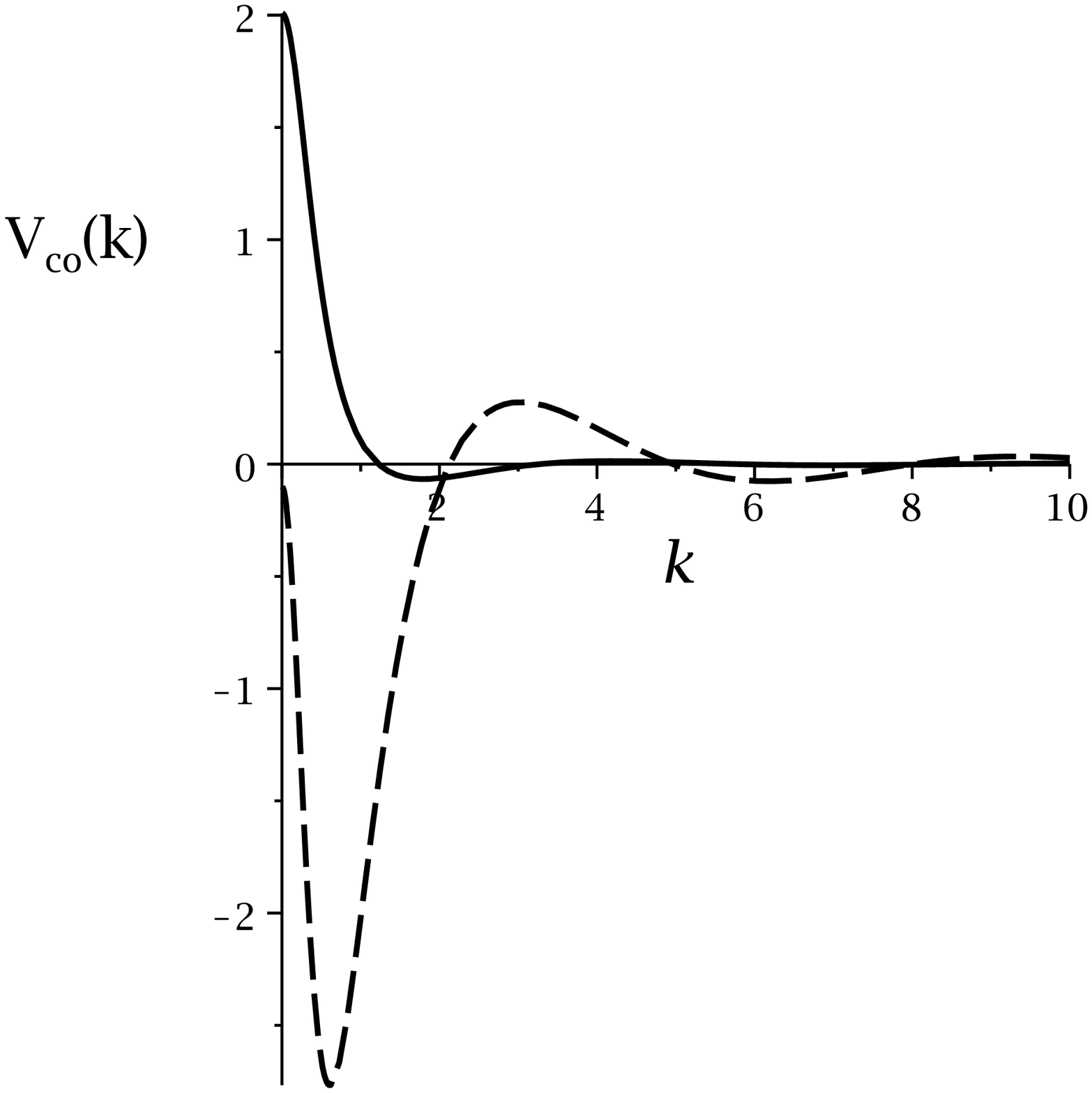}
\hfill
\includegraphics[scale=0.32]{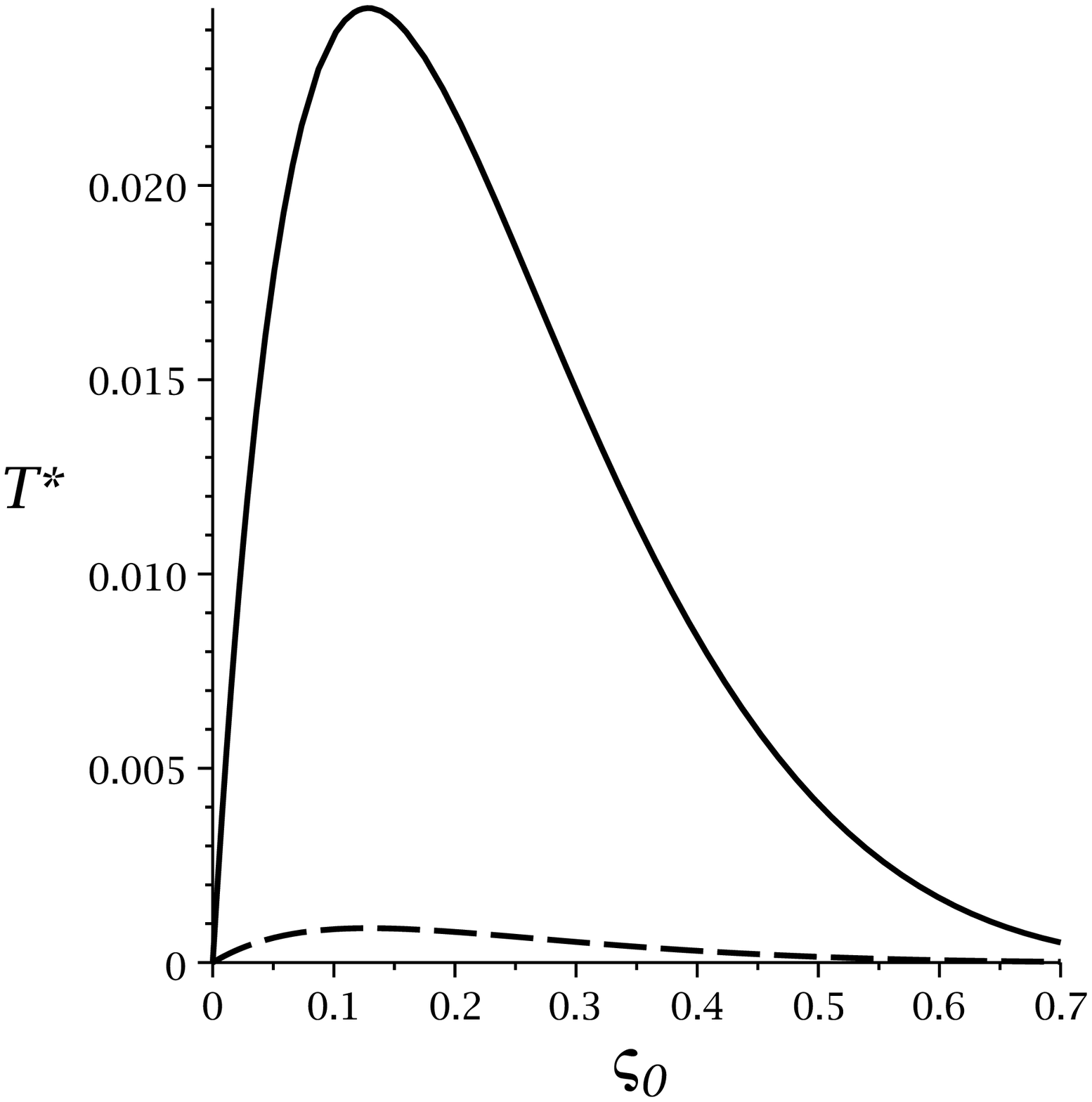}\hspace{5mm}}
\centerline{
\parbox[t]{0.48\textwidth}{\caption{ The potential $\tilde V_\mathrm{co}(k)$ for System 1 (solid line)
and for System 2 (dashed line). $\tilde V_\mathrm{co}(k)$ is in
dimensionless units, $k$ is in units $\sigma^{-1}$, where $\sigma$
is the particle diameter. \label{fig2}}}
\hfill
\parbox[t]{0.48\textwidth}{\caption{ The universal structural line (solid) in reduced units (see (\ref{T*})) and the metastable
 spinodal line of the separation into dilute and dense phases (dashed) for System 2.
Note that the temperature scale is different from the  corresponding scale
in the theory based on $\rho^*$~\cite{ciach:10:1}, and the
 scaling factor is $(6/\pi)^2$.
\label{fig3}}}
}
\end{figure}

We  choose two sets of parameters, considered in
reference~\cite{ciach:10:1} in the context of most probable inhomogeneous
structures
\begin{eqnarray}
\label{parameters}
\text{System}~1: \qquad {\cal A}_1=1, \quad {\cal A}_2=0.05, \qquad z_1=3, \quad z_2=0.5;
\nonumber  \\
\text{System}~2: \qquad {\cal A}_1=1, \quad {\cal A}_2=0.2,\ \ \, \qquad z_1=1, \quad z_2=0.5.
\end{eqnarray}
The relevant parameters, $k_\mathrm{b}$, $v_2^*$ and $v_0^*$, (see
(\ref{v2*}) and (\ref{v0*})) take the values: $v_2^*\approx 3.02$
and
\begin{eqnarray}
\text{System}~1: \qquad   k_\mathrm{b}\approx 1.79,\phantom{81} \qquad v_0^*\approx -30.145;
\nonumber \\
\text{System}~2: \qquad   k_\mathrm{b}\approx 0.6089,\, \qquad v_0^*\approx 0.035.\phantom{-3\,}
\end{eqnarray}
The two potentials in Fourier representation are shown in figure~\ref{fig2}. In
the first  system small clusters are formed, since $2\pi /k_\mathrm{b}$ is
small. Moreover, $\tilde V_\mathrm{co}(0)>0$, and the clusters repel each
other. The gas-liquid separation is entirely suppressed due to the very short range of the attractive part of the potential. In the second system  large clusters are formed, and $\tilde
V_\mathrm{co}(0)<0$  (the clusters attract each other). Therefore, in MF, the
metastable separation into disordered low- and high density phases
occurs at low temperature. Simulation results show that when large clusters are formed, gas-liquid separation occurs for low temperatures, and periodic phases are stable at higher temperatures~\cite{archer:07:0}.

We are interested mainly in the part of the phase diagram where the
homogeneous structure is less probable than periodic distribution of
particles in space, i.e. when $\Omega_\mathrm{co}$ does not assume a
minimum for $\zeta_0=\text{const}$ (see (\ref{p2})). We stress that the most probable structure differs from the average structure due to mesoscopic fluctuations. Cluster formation is associated with the excess volume fraction followed by a depleted volume fraction in mesoscopic regions, and the most probable mesoscopic structure associated with cluster formation is periodic. Displacements of the clusters (i.e., mesoscopic fluctuations) can destroy the long-range order, though. Indeed, when temperature is sufficiently high, the average volume fraction $\zeta({\bf r})$ takes the constant value as a result of the averaging over cluster displacements, and the disordered inhomogeneous structure with short-range correlations of the cluster positions is found~\cite{brazovskii:75:0,ciach:06:2}. On the other hand, for low temperatures, the ordered periodic structures are stable~\cite{brazovskii:75:0,ciach:06:2}. The phase-space region where the inhomogeneous phases (with either short- or long-range order) are stable is enclosed by the
structural line \cite{ciach:06:2,ciach:08:1} given by
$T^*=1/A_2(\zeta)$ and shown in figure~\ref{fig3}. The structural line is also
referred to as $\lambda$-line in literature~\cite{stell:99:0,
archer:07:0,archer:08:0,ciach:00:0,ciach:03:1}.  Note that in the
reduced units (see (\ref{T*})) the structural line is universal.

\begin{figure}[!h]
\centerline{\hspace{5mm}\includegraphics[scale=0.32]{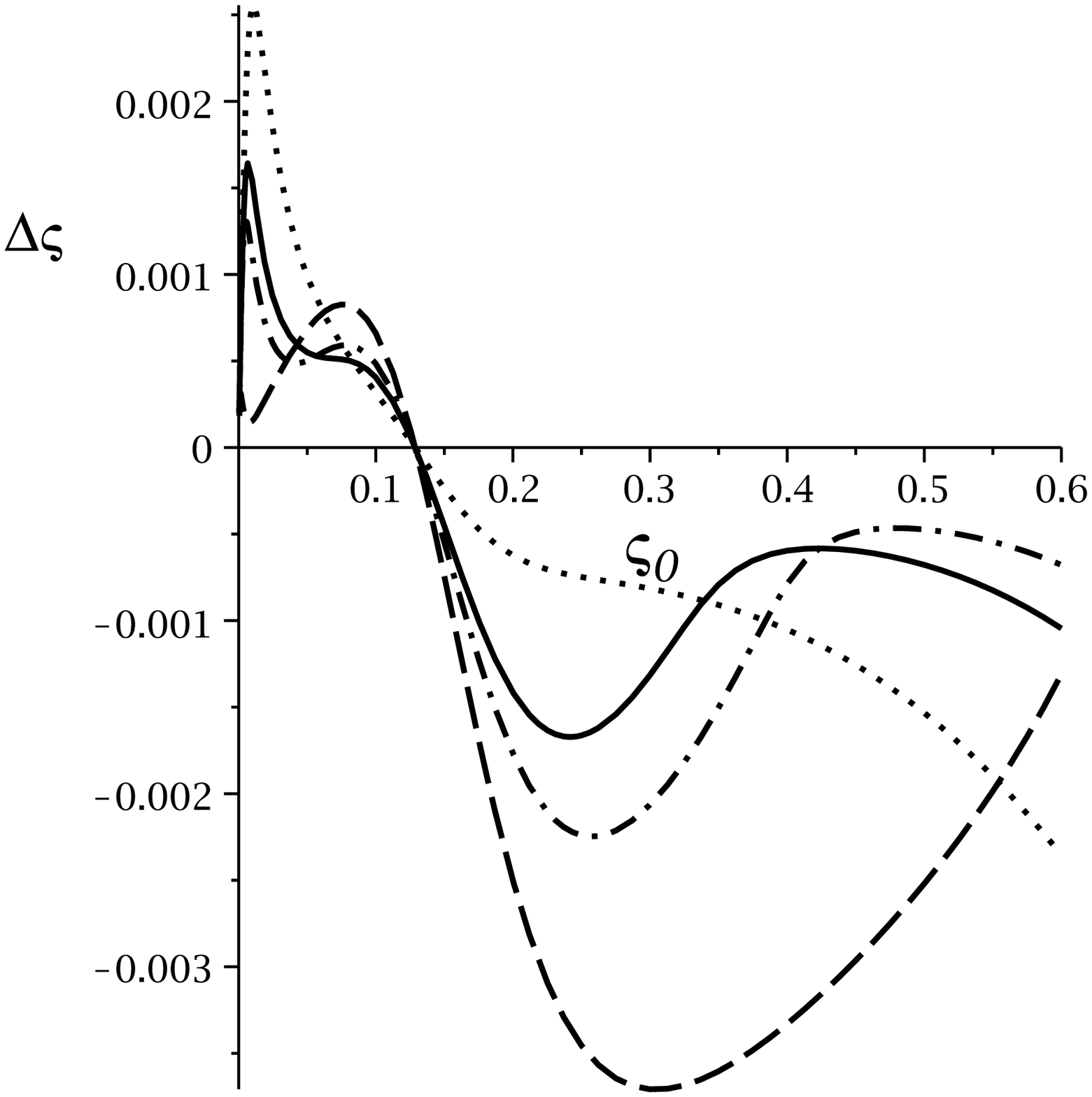}
\hfill
\includegraphics[scale=0.32]{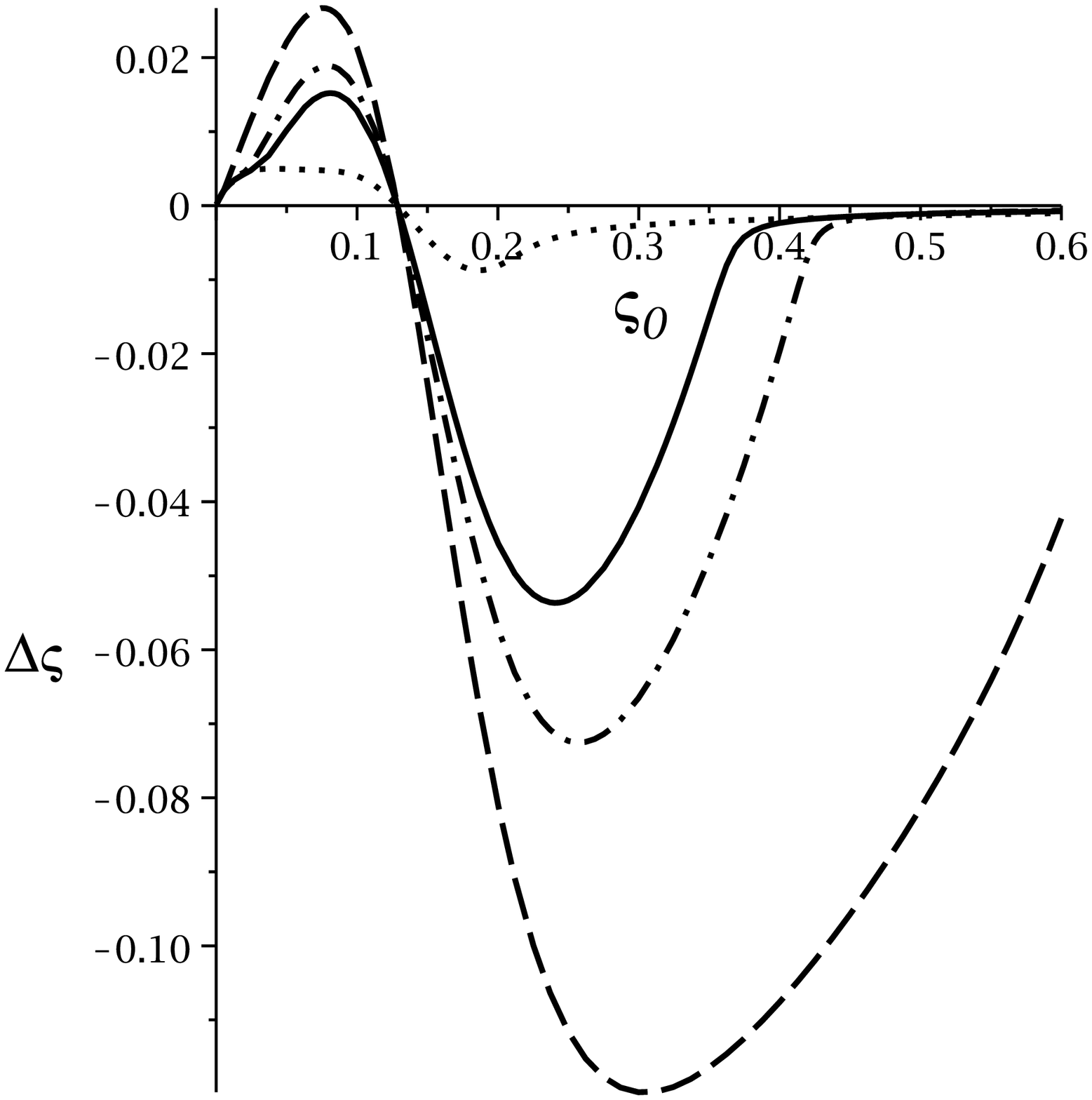}\hspace{5mm}}
\centerline{
\parbox[t]{0.48\textwidth}{\caption{ $\Delta\zeta(\zeta_0)$  for System~1.   Dotted, solid,  dash-dotted, and dash  lines correspond to
  $T^*=0.02,0.01, 0.007$ and $ 0.0008$, respectively.
equations~(\ref{shift2}) and (\ref{shift}) are not distinguishable on the plot.\label{fig4}}}
\hfill
\parbox[t]{0.48\textwidth}{\caption{ $\Delta\zeta(\zeta_0)$ for System~2 from equation~(\ref{shift2}).  Dotted, solid, dash-dotted  and dash
lines correspond to $T^*=0.02,0.01,0.007,0.0008$, respectively.
\label{fig5}}}}
\end{figure}
%
\begin{wrapfigure}{i}{0.5\textwidth}
\centerline{
\includegraphics[scale=0.315]{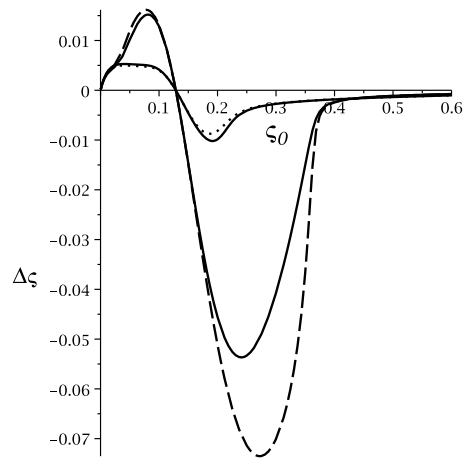}}
\caption{ Comparison of the two expressions, equations~(\ref{shift2}) and (\ref{shift}),  for $\Delta\zeta(\zeta_0)$ for System~2. The  lines with
smaller $|\Delta\zeta|$ correspond to  $T^*=0.02$; the solid line represents equation~(\ref{shift}), and the dotted  line represents equation~(\ref{shift2}).
The remaining lines represent
 equation~(\ref{shift}) (solid) and equation~(\ref{shift2}) (dashed) for $ T^*=0.01$.
\label{fig6}}
\vspace{-5mm}
\end{wrapfigure}

We first compare the change of the average volume fraction induced
by mesoscopic fluctuations. The  shift $\Delta\zeta$ calculated
from~(\ref{shift2}) and (\ref{shift}) in System~1 is shown in
figure~\ref{fig4}. The shift is small for a relevant range of temperatures,
and both formulas yield practically the same result~--- they are
indistinguishable on the plot. The shift increases for a decreasing
temperature. In System~2, the fluctuation contribution to the volume fraction is
much larger than in System~1 (figure~\ref{fig5}). As expected, when
$\Delta\zeta$ is not very small, $\Delta\zeta/\zeta_0\approx 0.2$,
then the two approaches yield somewhat different results, as shown
in figure~\ref{fig6}  for System~2.

\begin{figure}
\centerline{
\hspace{5mm}
\includegraphics[scale=0.32]{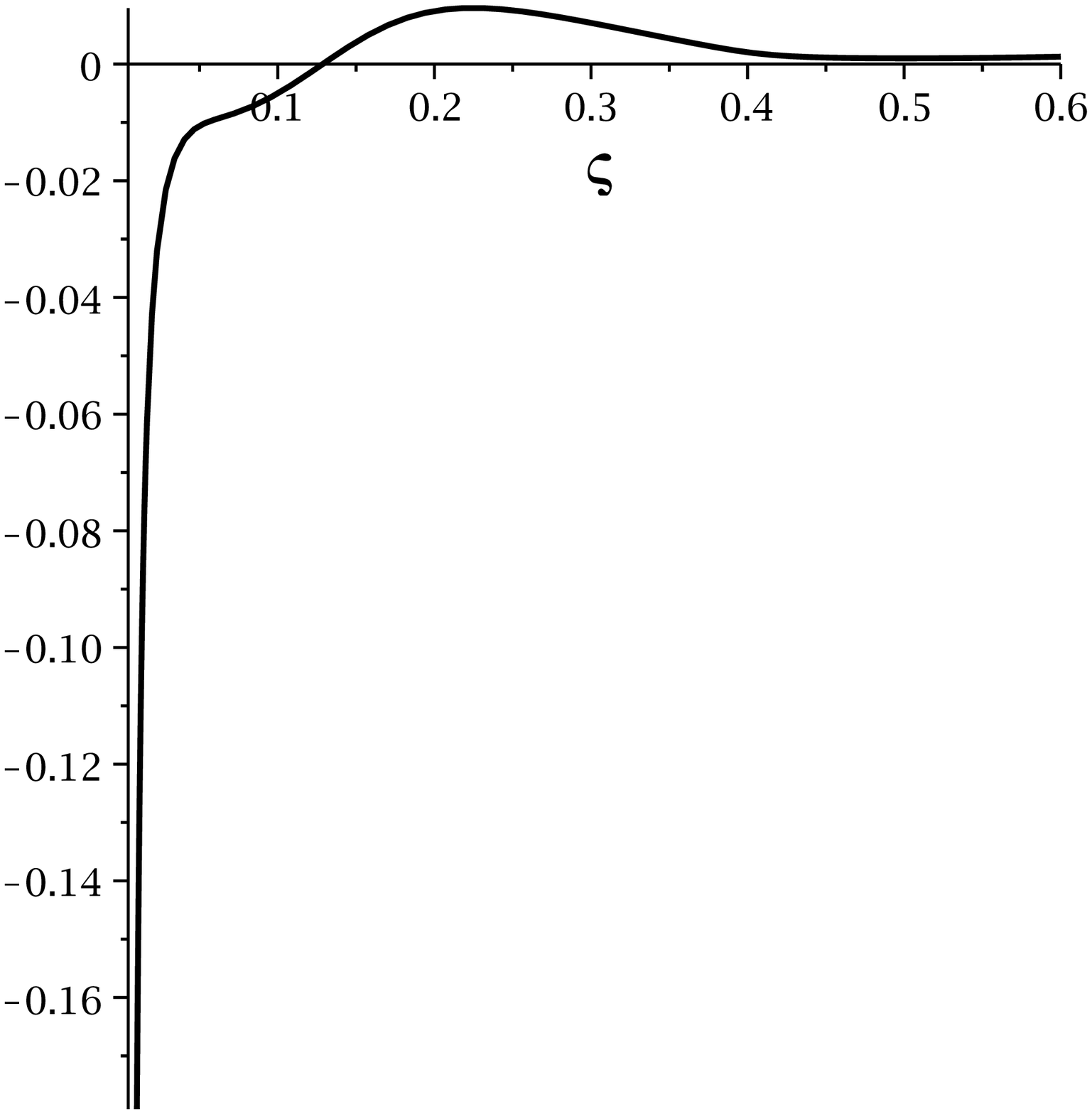}%
\hfill
\includegraphics[scale=0.32]{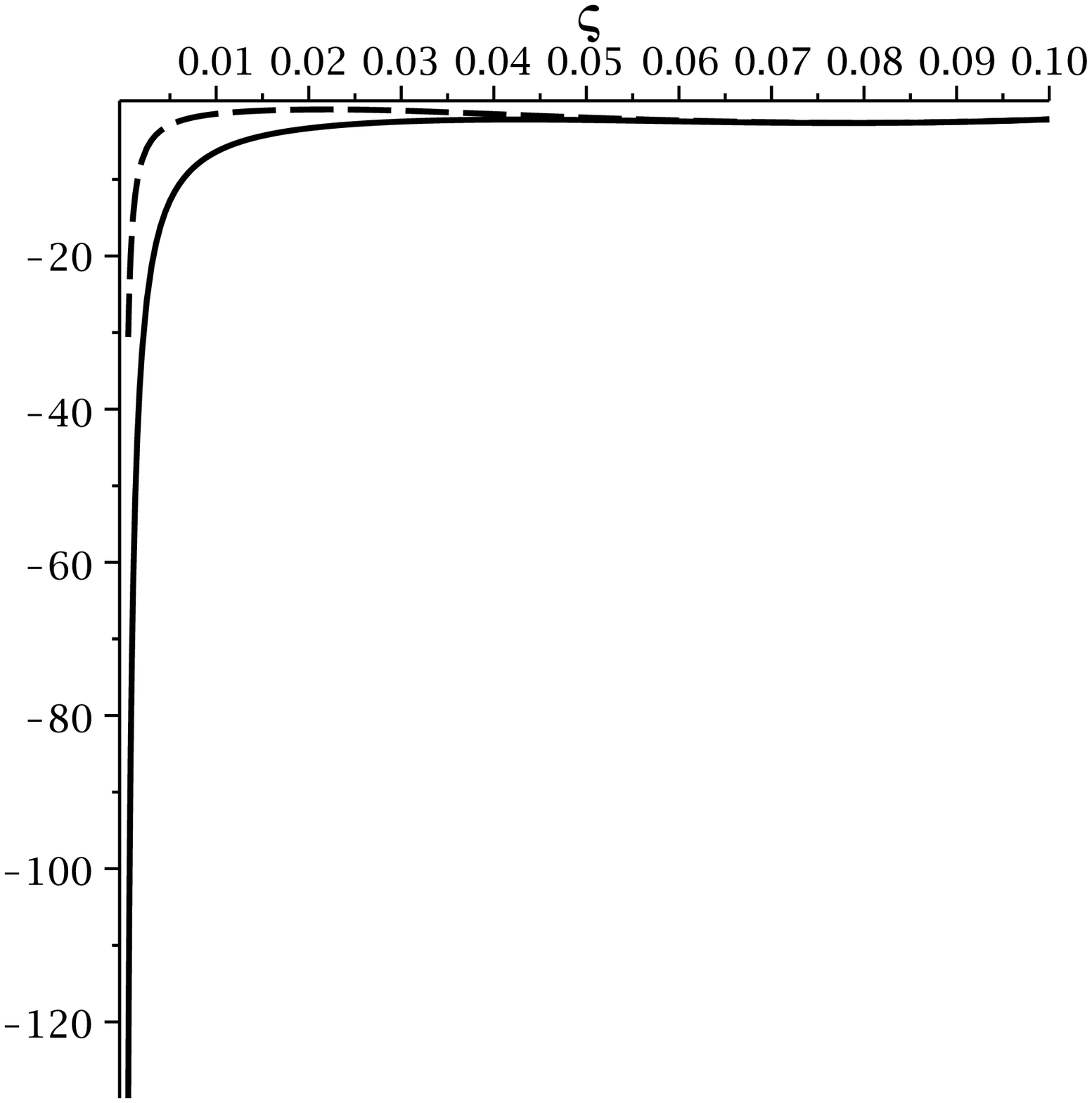}\hspace{5mm}}
\caption{Left panel: the relative difference,
$(\bar\mu-\bar\mu^\mathrm{MF})/\bar\mu^\mathrm{MF}$, between $\bar\mu$ calculated
from equation~(\ref{barmu}), and the MF approximation $\bar\mu^\mathrm{MF}$
(\ref{barmuMF})  for System~1 at $T^*=0.007$ and $\zeta\geqslant 0.01$.
Right panel: $\bar\mu-\bar\mu^\mathrm{MF}$ for System 1 (solid line) and
System 2 (dashed line) for $0.0005\leqslant\zeta\leqslant 0.1$. \label{fig7}}
\end{figure}

\begin{figure}[!b]
\centerline{\hspace{5mm}\includegraphics[scale=0.32]{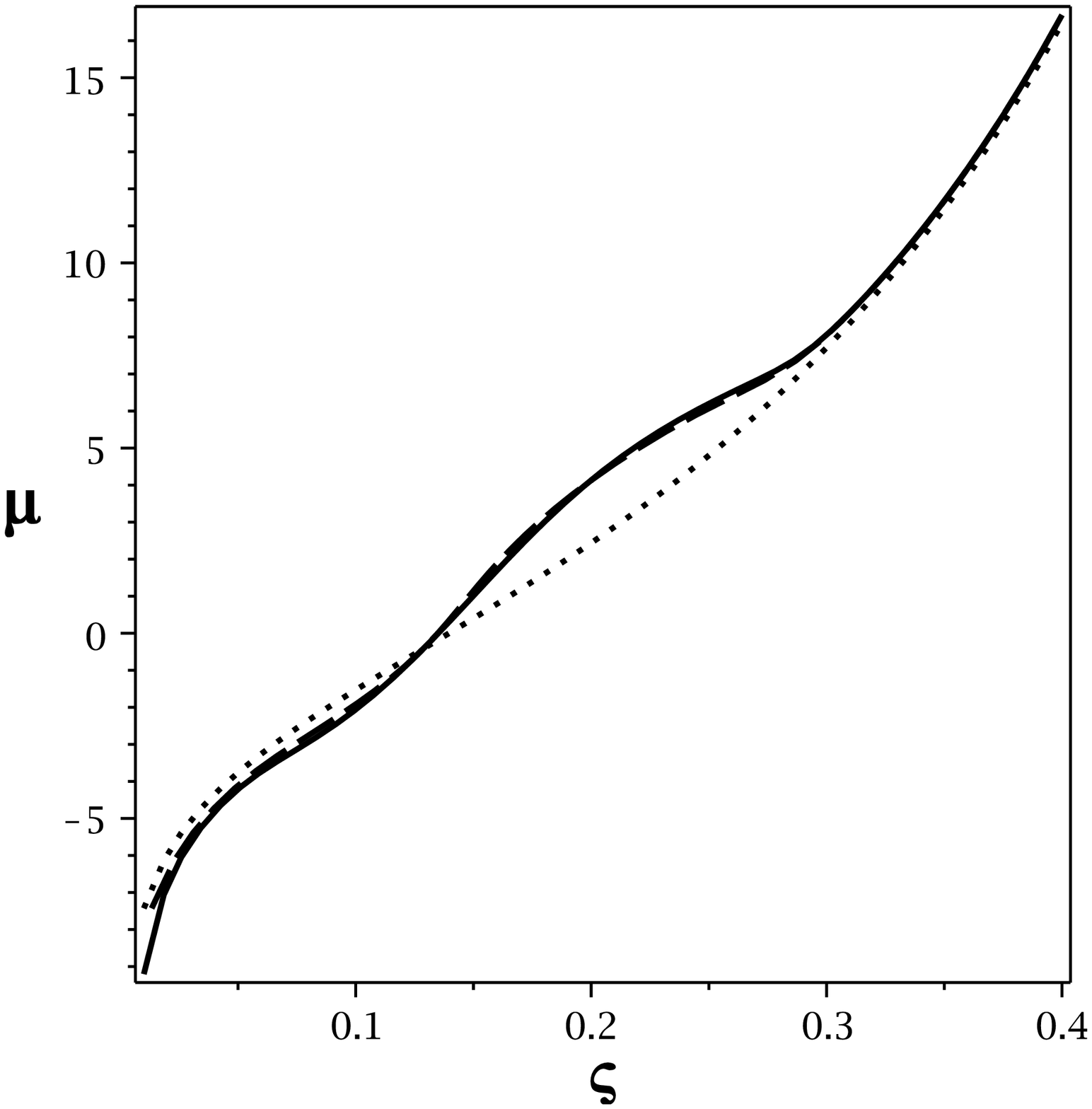}
\hfill
\includegraphics[scale=0.32]{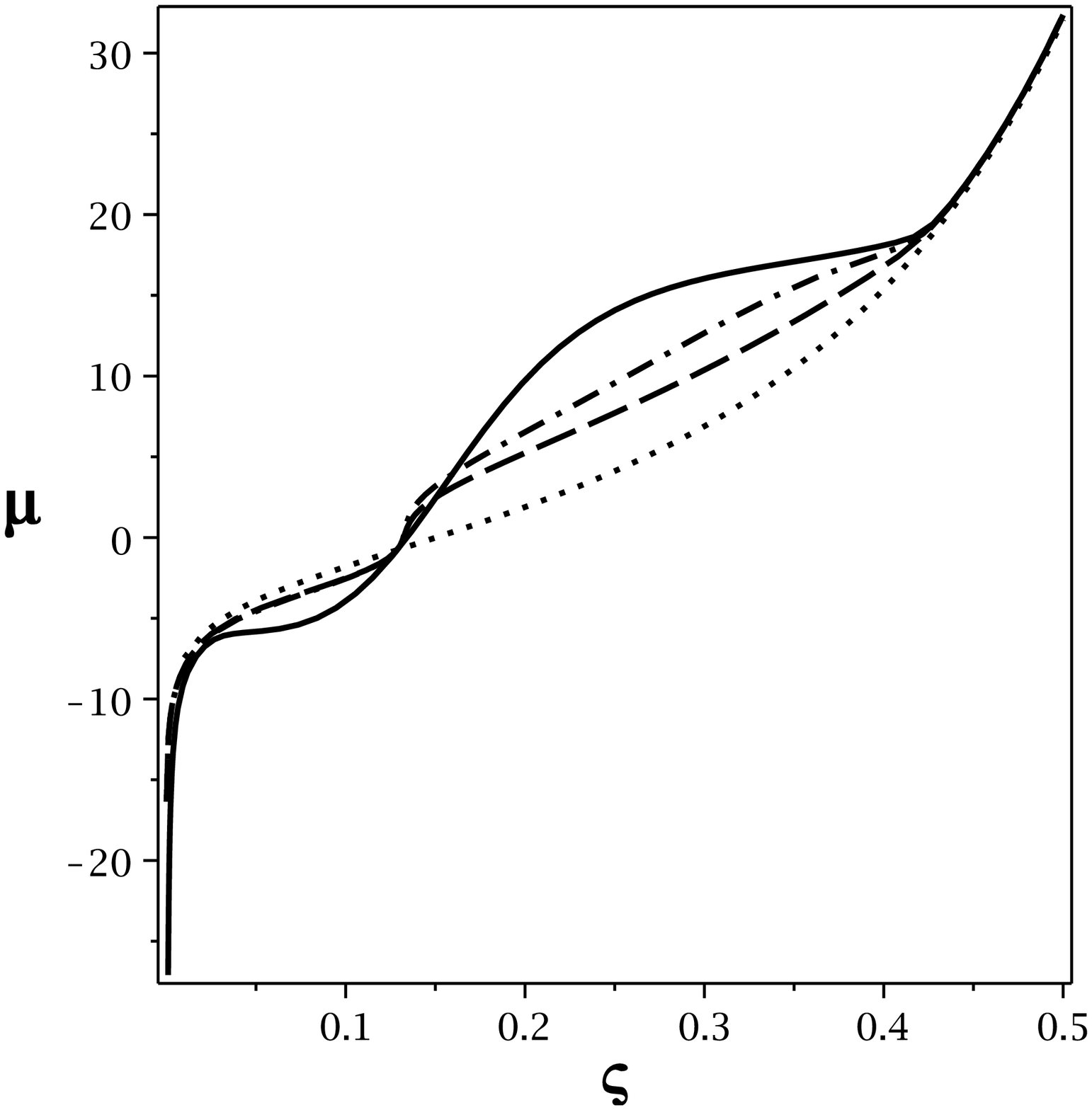}\hspace{5mm}}
\centerline{
\parbox[t]{0.48\textwidth}{\caption{Chemical potential in MF (dotted line), and with the fluctuation contribution included according to equation~(\ref{barmu})
 (solid)  and equation~(\ref{barmu0}) with (\ref{shift2}) (dashed line) for $T^*=0.015$ in System~2. The volume fraction is dimensionless,
and the chemical potential is in $\pi kT/6$ units.
\label{fig8}}}
\hfill
\parbox[t]{0.48\textwidth}{\caption{Chemical potential in MF (dotted line), and with the fluctuation contribution included according to equation~(\ref{barmu})
(solid line), equation~(\ref{barmu0}) with equation~(\ref{shift2}) (dashed line),  and equation~(\ref{barmu0}) with equation~(\ref{shift}) (dashed-dotted line) for $T^*=0.007$ in System~2.
The volume fraction is dimensionless, and the chemical potential is in $\pi kT/6$ units.
\label{fig9}}}}
\end{figure}

In the next step, we study the chemical potential. The fluctuation
contribution in System~1 is small, except at very small volume
fractions (figure~\ref{fig7}), whereas in System~2 it is substantial, and
increases for decreasing temperature, as shown in figures~\ref{fig8}   and~\ref{fig9}.
The two approaches yield similar results for small $\Delta\zeta$,
whereas when $\Delta\zeta/\zeta_0> 0.25$, significant discrepancy
between the two approaches is obtained. We can conclude that on the
quantitative level the approximate theory is oversimplified for the
range of $T$ and $\bar\zeta$ for which there are significant
discrepancies between the two approaches.

Note that since $\bar\mu-\bar\mu^\mathrm{MF}$ is large for volume fractions
$\zeta<0.03$, for very small volume fractions our results are
oversimplified.

\begin{figure}[!h]
\centerline{\hspace{5mm}\includegraphics[scale=0.32]{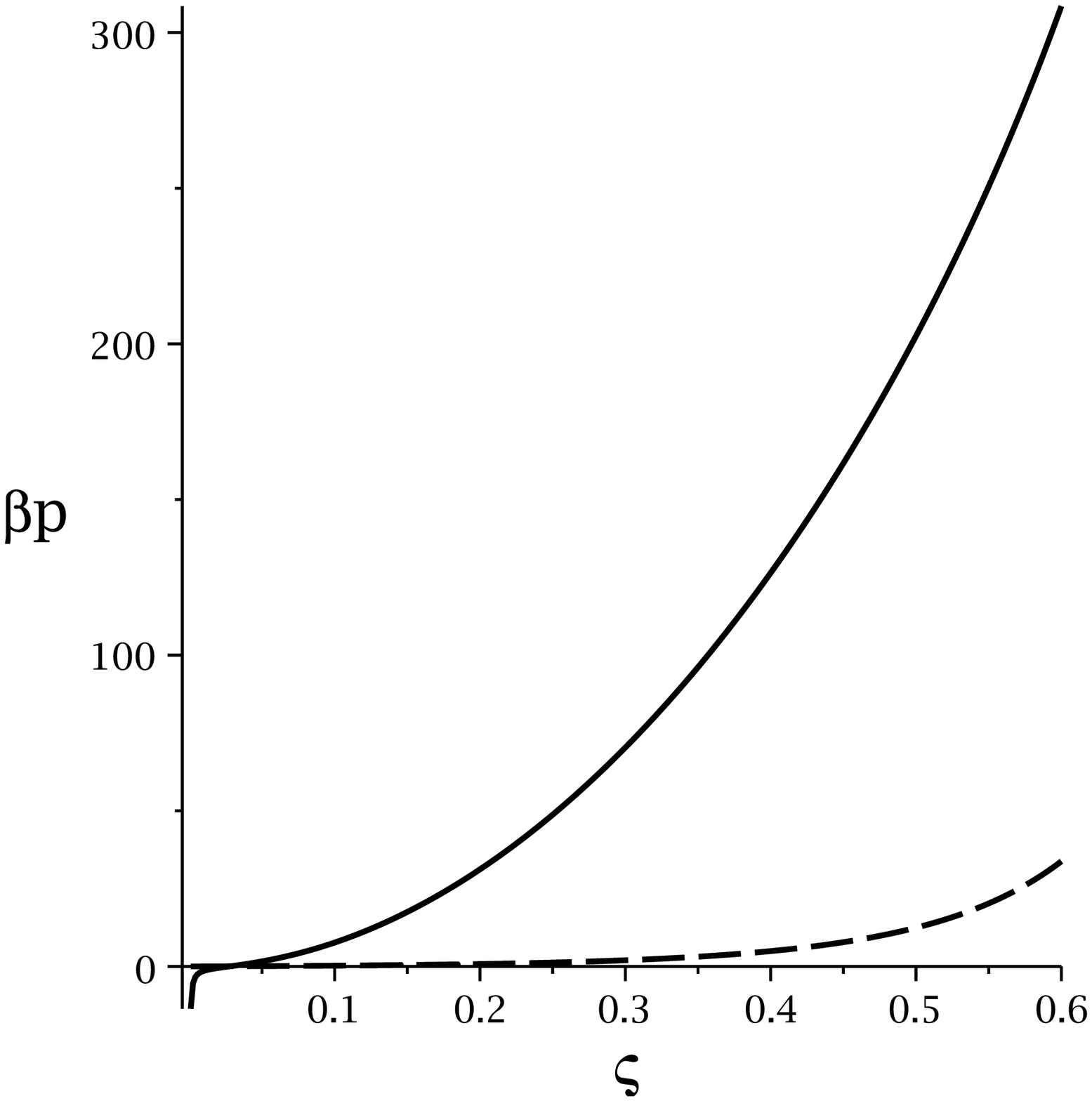}
\hfill
\includegraphics[scale=0.32]{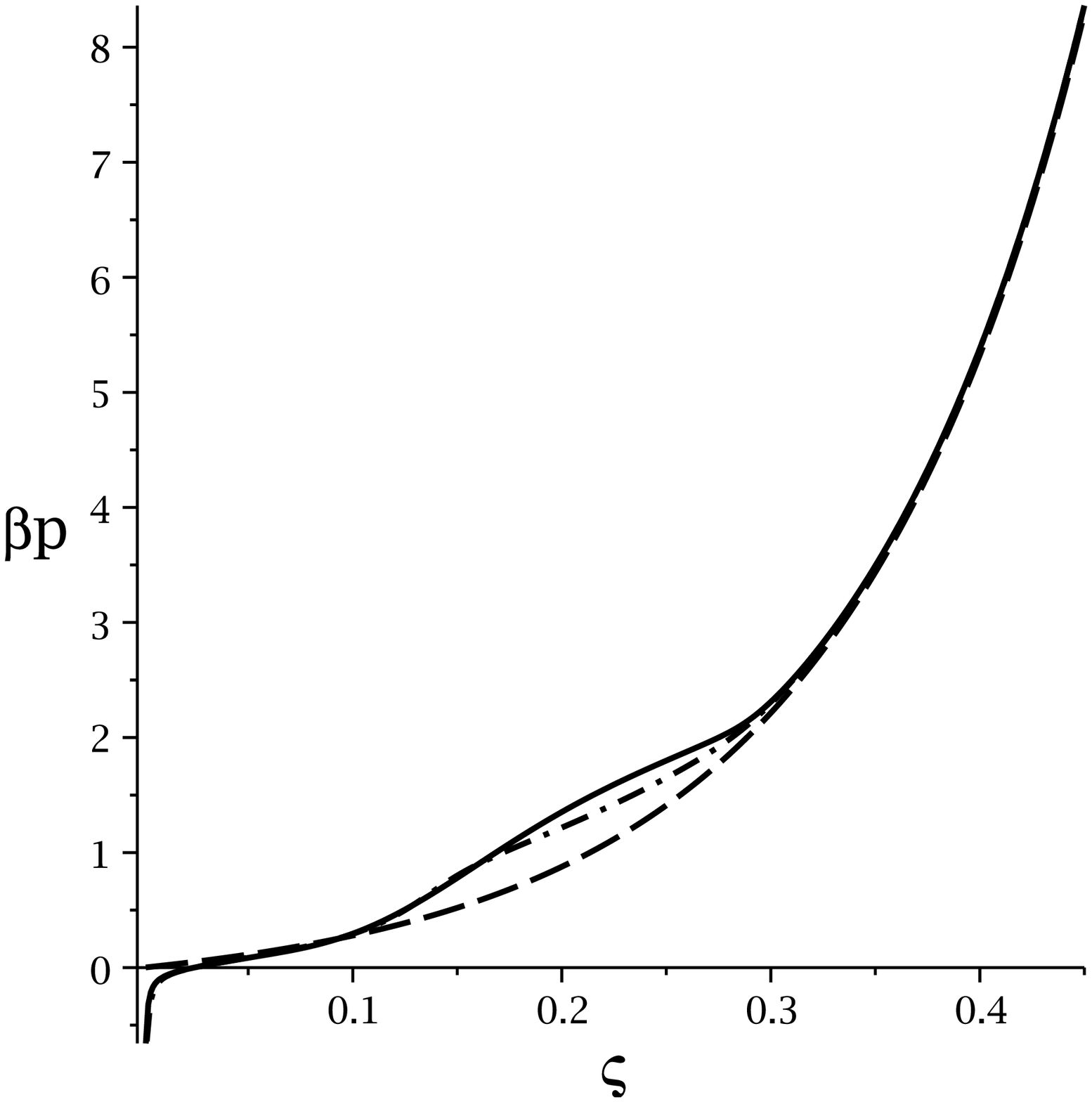}\hspace{5mm}}
\centerline{
\parbox[t]{0.48\textwidth}{\caption{ The EOS isotherms for System~1 for $T^*=0.02$. Dashed line is the MF result (equation~(\ref{EOSMF})) and
solid line represent equation~(\ref{EOS}), and equation~(\ref{EOS0}) with equation~(\ref{shift2}), indistinguishable on the plot.
\label{fig10}}}
\hfill
\parbox[t]{0.48\textwidth}{\caption{ The EOS  isotherms (\ref{EOS}) (solid line)  and (\ref{EOS0}) with equation~(\ref{shift2}) (dash-dotted line), and the MF approximation  (\ref{EOSMF})
(dashed line) for System~ 2 for $T^*=0.015$.
\label{fig11}}}}
\end{figure}

\begin{figure}[!b]
\centering
\hspace{5mm}
\includegraphics[scale=0.32]{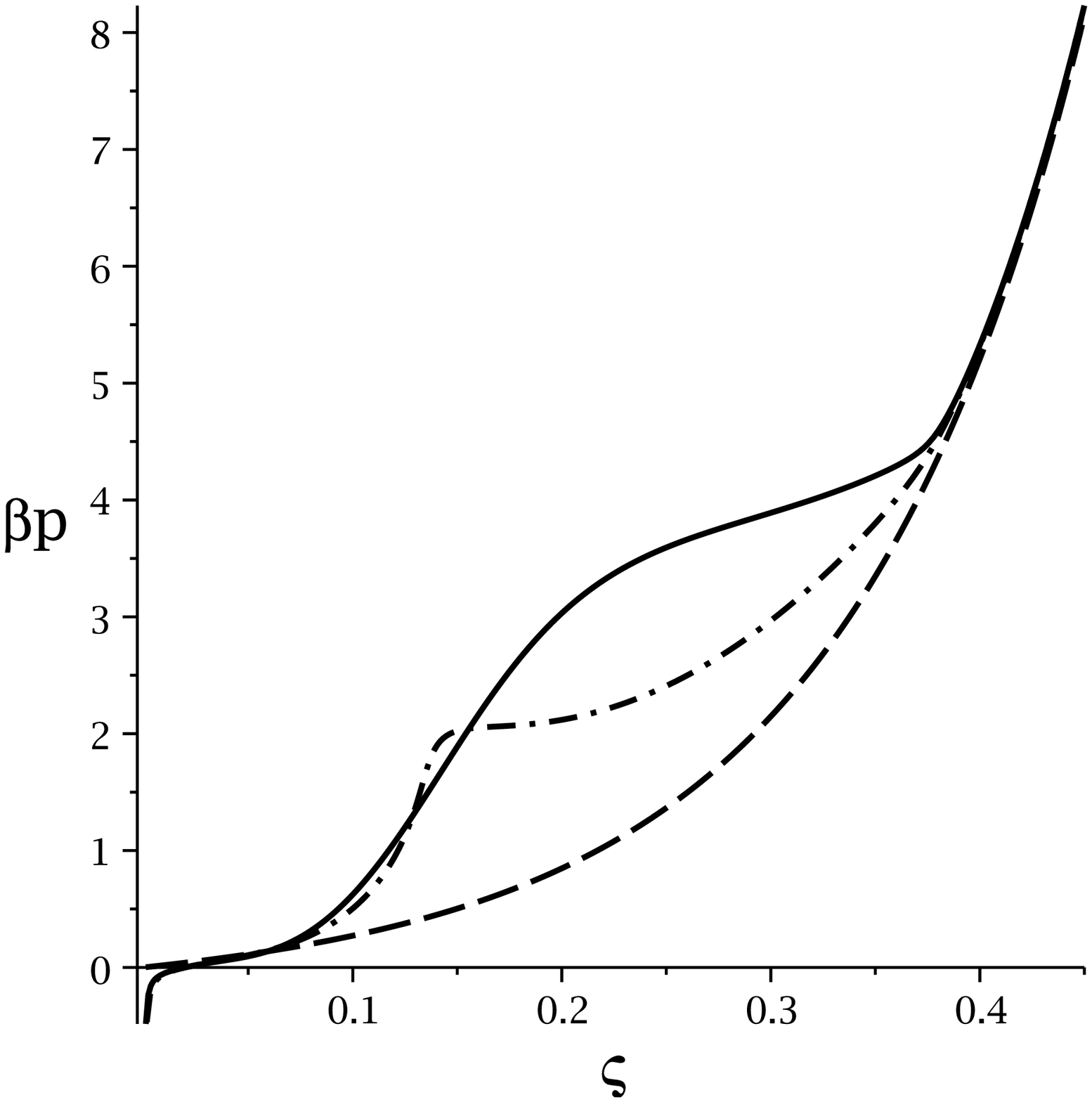}
\hfill
\includegraphics[scale=0.32]{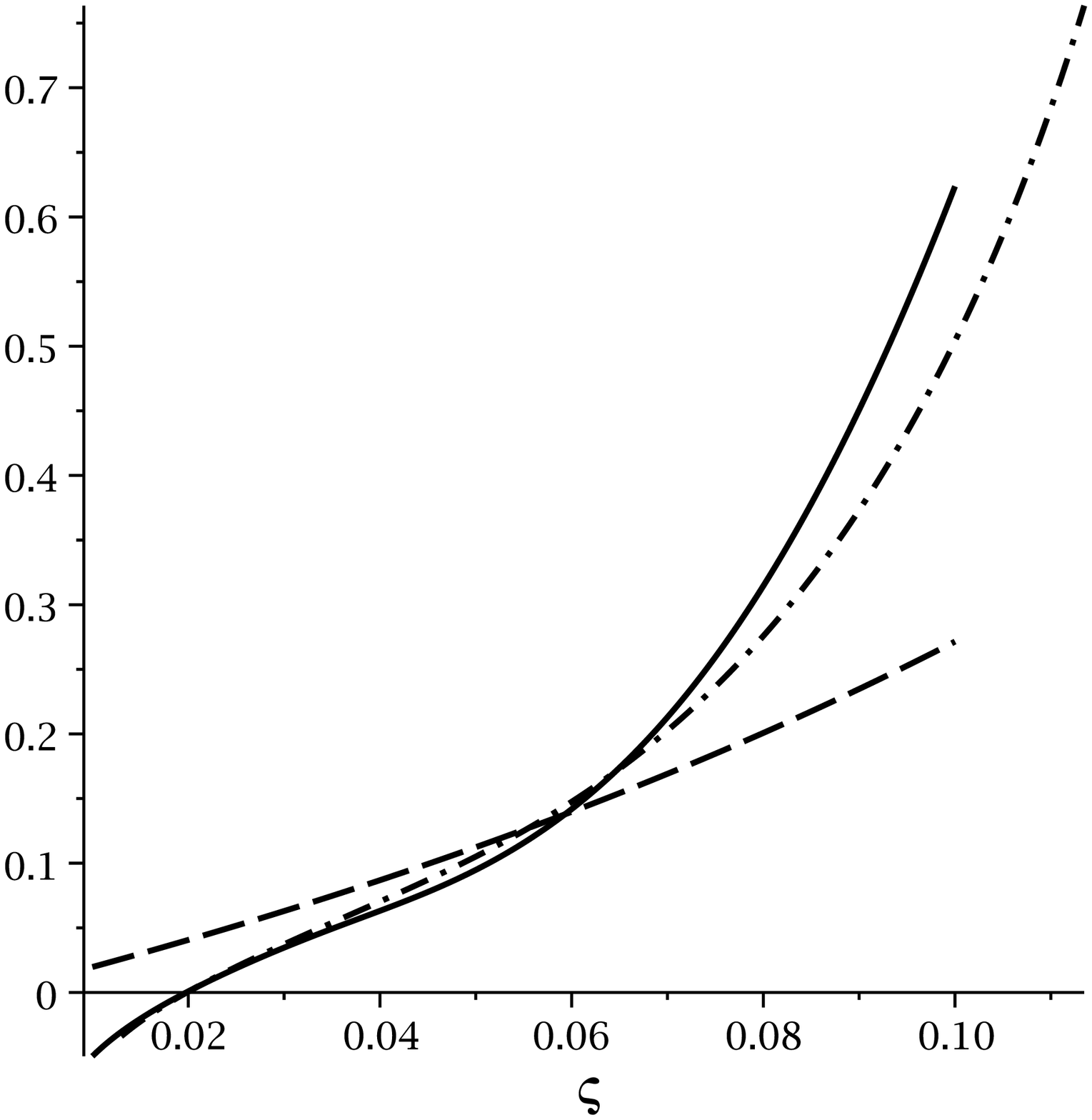}
\hspace{5mm}
\caption{ The EOS  isotherms (\ref{EOS})  (solid line)  and (\ref{EOS0}) with equation~(\ref{shift2}) (dash-dotted line),
 and the MF approximation
(\ref{EOSMF}) (dashed line) for System 2 for $T^*=0.0093$.
Left and right panels show large and small range of volume fractions, respectively.
\label{fig12}}
\end{figure}

\begin{figure}[!h]
\centerline{\hspace{5mm}\includegraphics[scale=0.32]{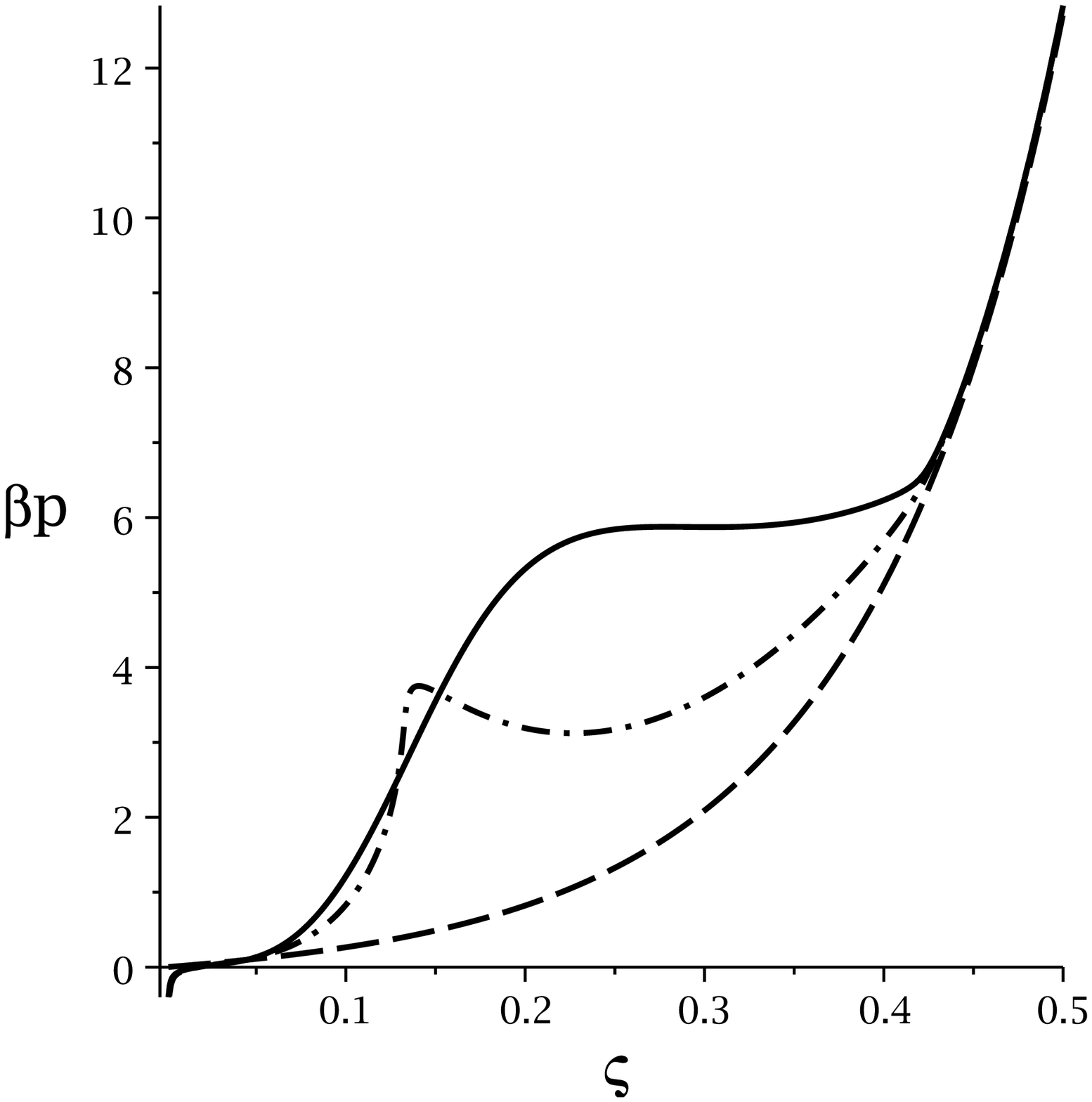}
\hfill
\includegraphics[scale=0.32]{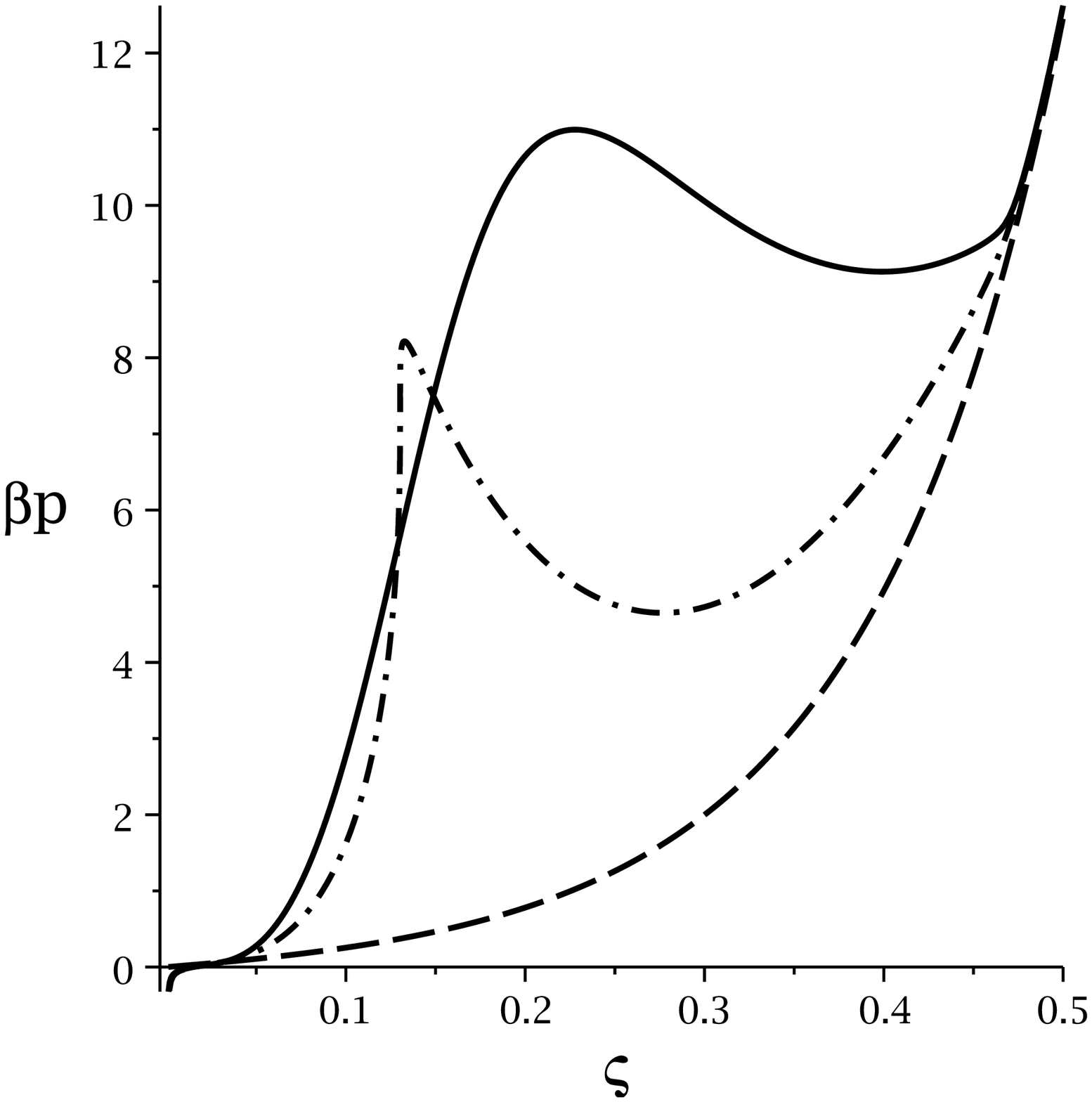}\hspace{5mm}}
\centerline{
\parbox[t]{0.48\textwidth}{\caption{ The EOS  isotherms (\ref{EOS}) (solid line)  and (\ref{EOS0}) with equation~(\ref{shift2}) (dash-dotted line), and the MF approximation  (\ref{EOSMF})
(dashed line) for System~2 for $T^*=0.007$.
\label{fig13}}}
\hfill
\parbox[t]{0.48\textwidth}{\caption{ The EOS  isotherms (\ref{EOS})  (solid line)  and (\ref{EOS0}) with equation~(\ref{shift2}) (dash-dotted line), and the MF approximation  (\ref{EOSMF}) (dashed line)
for System~2 for $T^*=0.005$. \label{fig14}}}}
\end{figure}

Finally, we present the isotherms obtained from (\ref{EOS}) and
(\ref{EOS0}) for the two systems in figures~\ref{fig10}--\ref{fig14}. In System~1 the
pressure is much higher than found in MF, and for all temperatures
it monotonously increases with $\bar\zeta$, as shown in figure~\ref{fig10}. The
increased pressure associated with mesoscopic fluctuations may
result from the repulsion between the clusters, because in this case
$\tilde V_\mathrm{co}(0)>0$.


Since in System~2 $\tilde V_\mathrm{co}(0)<0$, a  mechanical instability
develops at the MF spinodal line, with the metastable MF critical
point $T^*_\mathrm{c}\approx 0.0009$. What is really interesting is that such instability
appears at much higher temperatures due to mesoscopic
fluctuations. This is in strong contrast to the fluids with purely attractive interactions, where  density fluctuations decrease the critical temperature with respect to the mean-field estimate. The present case with dominant fluctuations associated with mesoscopic wavelengths bears some resemblance to the restricted primitive model (RPM) of ionic systems. There is no gas-liquid instability in the RPM at the MF level of a mesoscopic theory analogous to the one considered here, but when the short-wavelength charge-density fluctuations are included, such instability appears~\cite{ciach:00:0}. One could imagine that the mesoscopic fluctuations, i.e., displacements of the clusters from their most probable locations lead to their coalescence when $\tilde V_\mathrm{co}(0)<0$, and thus
support the phase separation.
 Note that properties of System~2 are completely
different from the previously studied System~1.

In this region of the phase diagram, the shift of the volume fraction
is large. Therefore, on the quantitative level, the results are not
sufficiently accurate. The inflection point on the $P(\bar\zeta)$
isotherm appears at $T_\mathrm{c}^*\approx 0.0093$ or $T^*_\mathrm{c}=0.007$ according
to~(\ref{EOS0}) or (\ref{EOS}) with (\ref{shift2}), respectively.
Both temperatures, however, are much higher than in MF. Further
studies are required to verify if the separation into disordered
inhomogeneous phases, or periodic ordering of clusters  occurs. If the ordered phases are formed, the still open question is for which
part of the phase diagram such phases are  globally stable.

Finally, let us focus on the pressure for very small volume fractions.
In the fluctuation correction to pressure (equation~(\ref{F}))
 the first term comes from the fluctuation contribution to chemical potential (see (\ref{barmu})).
As shown in figure~\ref{fig7} (right panel), for very low volume fractions, our
approximation is oversimplified, so the negative pressure is an
artifact. For very low volume fractions, we should expect a perfect gas
behavior, except that some fraction of particles should form
clusters. Pressure should be proportional to the sum of the number
densities of monomers and clusters. Since the number of clusters is
smaller than the number of particles forming them, pressure
should be
 smaller than in the corresponding perfect gas of isolated particles. Our theory agrees with this expectation
(see figure~\ref{fig12}, right panel).

\section{Summary}

In this work, the effects of mesoscopic fluctuations on the average volume fraction, chemical potential and pressure as functions of temperature and the average
volume fraction were considered within the framework of the mesoscopic theory~\cite{ciach:08:1,ciach:11:0}. We restricted our attention to a stable or metastable
disordered phase. The fluctuation contribution to the quantities mentioned above was calculated in two ways. First, we considered the fluctuations about the average volume
fraction, and  derived equations~(\ref{shift}), (\ref{barmu}) and (\ref{EOS}) for the volume fraction, chemical potential and pressure, respectively. In the second version,
we considered the fluctuations about the most probable volume fraction, and obtained equations~(\ref{shift2}), (\ref{barmu0}) and (\ref{EOS0}), with $\zeta_0$ satisfying
(\ref{avdg0}). The chemical potential and the EOS as functions of the average volume fraction are given by parametric equations~(\ref{shift2}) and (\ref{barmu0}),
and (\ref{EOS0}), respectively.
Our expressions are derived under the assumption that the dominant fluctuations are of small amplitudes. Consistent with the above assumption,
the two methods yield the same result to a linear order in  the fluctuation contribution to the volume fraction $\Delta\zeta$.

The fluctuation contributions to all three quantities  were explicitly calculated for two versions of the SALR potential. In System~1 the zeroth moment of
the effective interactions is positive and small clusters are formed. In System 2  the zeroth moment of the effective interactions is negative, and the clusters
are large. We obtain nearly the same results independently of the method used when the fluctuation-induced shift of the volume fraction is very small (System~1).
When $\Delta\zeta/\zeta_0\sim 0.2$ (low temperature in System~2), significant discrepancies between the  two methods appear for some part of the phase diagram.
The largest discrepancies are present when both methods yield the results that strongly deviate from the MF predictions.
The larger
is the probability of finding inhomogeneous mesoscopic states compared to the homogeneous distribution of particles,  the stronger are
the discrepancies between the two methods. It is in this part of the phase diagram
that the periodic order may appear. We conclude that the first method is superior to the second one because it is easier to implement. Exact results would
be necessary to get a comparison between the accuracy achieved by these methods.

We have found that mesoscopic fluctuations play a very important role and lead to a significant change of the chemical potential and pressure.
The larger is the probability of finding the inhomogeneities, i.e., the further away from the structural line on the low-$T$ side of it (figure~\ref{fig2}), the larger is the role of fluctuations. When
small clusters are formed (System~1 in section~4), the fluctuation contribution to pressure increases monotonously with an increasing volume fraction. By contrast,
for large clusters (System~2 in section~4) the fluctuation contribution to pressure is nonmonotonous; it is negligible for small as well as for large volume fractions,
whereas for intermediate volume fractions it is large and increases with a decreasing temperature (figures~\ref{fig10}--\ref{fig13}). Moreover, an  inflection point at the pressure -- volume
fraction isotherm appears at the  temperature and volume fraction  both much larger than found in MF. Further studies are
required to verify if the
periodically ordered cluster phases are stable in System~2, or phase separation occurs due to the mechanical
instability. Possible scenarios are:
(i) phase separation at low $T$, and periodic structures at higher $T$,
or (ii) the phase separation is only metastable, and finally (iii) the periodically ordered   phases are only metastable.
For System 1, the phase separation is not expected.

In addition to the assumptions discussed earlier, we make an approximation concerning the form of the effective potential
$V_\mathrm{co}$. Note that in equation~(\ref{int_pot_r}) we assumed that the pair distribution function vanishes for $r<1$ in $\sigma$
units. For the volume fraction, this is a poor approximation, and quantitative results for the structural line depend on the
form of the pair distribution function (regularization of the potential~\cite{patsahan_meryglod:04}).

In the future studies, the EOS for periodically ordered cluster
phases should be determined in order to find the phase diagram. Our
results indicate that despite the universal properties of the
dependence of the most probable structures on the thermodynamic
state, the effect of fluctuations on the average distribution of
particles may depend on the form of the interaction potential,
especially on the sign of the zeroth moment of the effective
interactions.

\section*{ Acknowledgements} A part of this work  was realized within the
International PhD Projects Programme of the Foundation for Polish
Science, co-financed from European Regional Development Fund within
Innovative Economy Operational Programme ``Grants for innovation''.
Partial support by the Ukrainian-Polish joint research project under
the Agreement on Scientific Collaboration between the Polish Academy
of Sciences and the National Academy of Sciences of Ukraine for
years 2009--2011 is also gratefully acknowledged.
\newpage

\newpage
\ukrainianpart

\title%
{Вплив мезоскопічних флуктуацій на  рівняння стану кластероутворювальних систем%
}
\author{А. Цях\refaddr{label1},
       О. Пацаган\refaddr{label2}}
\addresses{
\addr{label1} Інститут фізичної хімії, Польська академія наук, 01-224 Варшава, Польща
\addr{label2} Інститут фізики конденсованих систем Національної академії наук України, \\ вул. Свєнціцького, 1, 79011 Львів, Україна }

\makeukrtitle

\begin{abstract}
\tolerance=3000%
Рівняння стану для систем частинок, що самоскупчуються в агрегати, є
отримане в рамках мезоскопічної теорії, що поєднує метод функціоналу
густини і теоретико-польовий підхід. Ми досліджуємо вплив
мезоскопічних флуктуацій у невпорядкованій фазі. Явно обчислено
ізотерми `тиск -- об'ємна частка' для двох наборів параметрів потенціалу
короткосяжне притягання плюс далекосяжне відштовхування. В кожному
випадку врахування мезоскопічних флуктуацій приводить до підвищення
тиску, за винятком дуже малих об'ємних часток. Коли утворюються великі
кластери, механічна нестійкість системи присутня при набагато вищих
температурах, ніж це було отримано в наближенні середнього поля. В
цьому випадку фазове відокремлення конкурує із формуванням періодичних
фаз (колоїдних кристалів). У випадку малих кластерів механічна
нестійкість, пов'язана з відокремленням в розріджену і густу фази, не
виникає.

\keywords  кластери, самоскупчення, рівняння стану, мезоскопічні флуктуації%
\end{abstract}

\end{document}